\documentclass[%
 reprint,
 amsmath,amssymb,
 aps,
 longbibliography,
floatfix,
prx]{revtex4-2}

\usepackage{silence}
\usepackage{graphicx}
\usepackage{dcolumn}
\usepackage{bm}
\usepackage[colorlinks=true, linkcolor=blue, citecolor=blue, urlcolor=blue]{hyperref}

\usepackage{enumitem}
\usepackage{multirow}
\usepackage{algorithm}
\usepackage{algpseudocode}
\usepackage{makecell}
\usepackage{hhline}
\usepackage{adjustbox}
\usepackage{relsize}
\usepackage{float}
\usepackage{multirow}
\usepackage[T1]{fontenc}
\usepackage{longtable}
\usepackage{array}
\usepackage{tabularx}

\usepackage{xcolor}
\definecolor{YKB}{rgb}{0.00,0.20,0.75}
\definecolor{mygreen}{rgb}{0.00,0.65,0.05}

\usepackage[utf8]{inputenc}
\usepackage{mathptmx}
\usepackage{helvet} 
\usepackage{amsmath,amssymb}
\usepackage{siunitx}
\DeclareSIUnit\Molar{\textsc{m}}
\DeclareSIUnit\year{yr}

\begin{document}

\preprint{APS/123-QED}

\title{Interplay of Fidelity and Diversity in the Evolution of the Genetic Code}

\author{Yudam Seo}
\affiliation{%
 Department of Science Education, Seoul National University, Seoul 08826, Korea
}%

\author{Tsvi Tlusty}
\affiliation{%
Department of Physics, Ulsan National Institute of Science and Technology, Ulsan, 44919, Korea
}

\author{Junghyo Jo}%
 \email{jojunghyo@snu.ac.kr}
\affiliation{%
Department of Physics Education and Center for Theoretical Physics and Artificial Intelligence Institute, Seoul National University, Seoul 08826, Korea \\
}%
\affiliation{%
School of Computational Sciences, Korea Institute for Advanced Study, Seoul 02455, Korea \\
}%

\date{\today}

\begin{abstract}
The origin and organizing principles of the genetic code remain fundamental puzzles in life science. The vanishingly low probability of the natural codon-to-amino acid mapping arising by chance has spurred the hypothesis that its structure is a solution optimized for robustness against mutations and translational errors. For the construction of effective molecular machines, the dictionary of encoded amino acids must also be diverse enough in physicochemical features. Here, we examine whether the standard genetic code can be understood as a near-optimal solution balancing these two objectives: minimizing error load and aligning codon assignments with the naturally occurring amino acid composition. Using simulated annealing, we explore this trade-off  across a broad range of parameters. We find that the standard genetic code lies near local optima within the multidimensional parameter space. It is a highly effective solution that balances fidelity against resource availability constraints. These results suggest that the present genetic code reflects coevolution under conflicting pressures of fidelity and diversity, offering new insight into its emergence and evolution.
\end{abstract}

\maketitle

\section{Introduction: contingency, constraints, and optimization at the origin of the code}

The standard genetic code (SGC) is nearly universal, with only a few minor variations in the mapping of DNA codons to amino acids~\cite{Knight01,Osawa92}. With 64 codons encoding 20 amino acids and a stop signal, combinatorics allows for a staggering number of $21! \cdot S(64, 21) \sim \num{e84}$ possible mappings  -- all partitions of 64 codons into 21 distinct non-empty sets (or $ {\sim} \num{e79}$ excluding the stop codons~\cite{Blazej18, Schnauer97}). 
Thus, the problem of the SGC's origin is bifurcated into two interrelated questions. First, about \emph{stabilized universality}: why is one, nearly-universal map almost perfectly conserved throughout the tree of life? (unlike typical genetic traits that vary among individuals and populations). And a second question about the \emph{initial condition}: is there anything special about this particular map?  
  
Francis Crick’s {\it frozen accident theory} \cite{Crick68} addresses the first question:  the code's near-universality is a consequence of its function as a global dictionary, mapping 64 symbols (codons) to 20 meanings (amino acids) ~\cite{DiGiulio2005,Tlusty10}. Any perturbation of this dictionary after the emergence of complex life would propagate catastrophically across the proteome, imposing an intolerable evolutionary load. This mechanism dictated that the evolution of the code arrested—it ``froze''—very early in the history of life.
As for the second question, about the initial condition (the specific codon assignments), Crick merely proposed that this is a product of historical contingency, an "accident" that became fixed. 

The profound challenge to the pure `accidental' view lies in the SGC's conspicuous nonrandom structure~\cite{Sonneborn1965,Sella2006}. Particularly, codons differing by a single base (point mutation) are overwhelmingly assigned to amino acids with similar physicochemical properties~\cite{Woese66}. This is not the signature of chance, but rather the residue of selection or physical interactions. The freezing scenario has also been challenged, where some suggest that universality emerged in early communal life, as a result of selection for ``innovation-sharing protocols''~\cite{Vestigian2006}.  

The \emph{stereochemical theory} \cite{Gamow54,Dunnill1966,Yarus1989} suggests such a physicochemical origin, postulating a primitive, direct affinity between amino acids and their cognate (anti)codons. This implies that the code's mapping stems from intrinsic physical properties of matter. The predictive power of this theory, however, is somewhat limited by a lack of definitive experimental evidence for the requisite, explicit affinities. The continued function of tRNAs with artificially altered anticodons, along with the low, non-specific binding energies measured for RNA bases with amino acids, suggest that a purely stereochemical link did not dictate the final code.  Its contribution was likely an initial bias, setting the boundaries for what was chemically plausible in the earliest, non-enzymatic translation systems.   \cite{Rowe94,Knight1999,Yarus2005,Johnson2010}. 

\emph{The error minimization theory} is premised on the observation that the codon assignment of the SGC efficiently reduces the phenotypic cost of point mutations as well as errors occurring in transcription and translation processes \cite{Goldberg1966,Alff-Steinbeger69,Rumer2016trans,Tlusty2007,Massey08,Koonin09}.  In this view, the SGC is a highly optimized mapping that emerged by selection to maintain coherent biological information against the background noise of molecular imperfection. Computer simulations by Haig and Hurst on error minimization in the genetic code \cite{Haig91} revealed statistical differences related to codon positions. 
Further statistical analysis by Freeland and Hurst of the manifold of all possible genetic codes demonstrates the profound improbability of the SGC's specific configuration arising by chance \cite{Freeland98, Freeland98_2}. This study positioned the SGC at an extreme end of the distribution, with its superior error resilience estimated to be a statistical outlier with a probability of roughly ``one in a million'', which also reflects a bias in error positions (for further review of the error minimization theory, see~\cite{Ardell98, Freeland00,DiGiulio2005,Tlusty10}).

Error minimization has provided a quantitative lens to understand the order observed in the code. But it was also recognized that it is not the sole determinant of the mapping, but part of a tradeoff with other essential factors, particularly the \emph{diversity} of the amino acid vocabulary. At the extreme, a code singularly designed for error tolerance would constitute a degenerate state encoding a single amino acid, completely lacking the requisite \emph{coding capacity} to forge complex life. Such considerations led to the articulation of "physicochemical diversity" measures~\cite{Ardell2001} and the identification of ``diversifying steps'' along coevolutionary trajectories~\cite{Sella2006}. 

In previous works, we integrated the imperative for functional diversity within the information-theoretic framework of the SGC's emergence and evolution~\cite{Tlusty2007,Tlusty08_2,Tlusty10}, explicitly accounting for diversity in the code performance measure~\cite{Tlusty08,Tlusty2008c}. Yet, all these models ultimately relied on simplified, qualitative measures of diversity, and focused on generic symmetry-breaking aspects, while avoiding the subtleties of the present amino acid distribution.

The present work aims to resolve this ambiguity by rigorously quantifying diversity. Crucially, these estimates of diveristy are performed against a realistic, non-uniform statistical ensemble of the 20 amino acids. Earlier studies have included amino acid frequencies as weights in calculating the translation error~\cite{Gilis01, Tlusty08_2}. Here, we align the amino acid abundances in the SGC and their respective codon usage to estimate diversity, accounting for evolutionary lineage and genetic variation across all species sharing the universal dictionary of the SGC~\cite{Vestigian2006}.  

By demonstrating that the SGC is nearly optimized with respect to the empirical composition of the modern proteome, we provide new evidence supporting a scenario in which the physicochemical spectrum of the code is finely-tuned to match these material demands. Our analysis shows that the code's final, frozen architecture reflects the material constraints set by the current composition of molecular machines. This explains, for instance, the redundant encoding of highly utilized residues, such as leucine, serine, and arginine, which are allocated multiple codons to ensure the efficient mass production of cellular machinery.
This suggests that the SGC had achieved its terminal emergent optimality—balancing high throughput and accuracy within the physico-chemical constraints of current molecular machinery---just before its universal structure froze. 

 In Section~\ref{sec:theory}, we introduce variations in codon mutation rates and define the two objective terms of code optimality: (i) codon translation errors and (ii) amino acid compositional alignment. In Section~\ref{sec:results}, we show that the SGC is optimal under local variation and demonstrate that it can be achieved by balancing these two objectives. We then examine genetic codes across different species and show that they occupy exceptionally rare and optimal configurations compared to random codes. Finally, in Section~\ref{sec:discussion}, we discuss the implications and limitations of our study. Appendix~\ref{sec:appendix} provides an analytical derivation of the mutation weight.

\section{Measuring the genetic code's performance}
\label{sec:theory}

In the following, we construct a measure of code performance, taking into account error resilience and functional diversity.

\subsection{Codon mutation rate variation}
\label{subsec:weights}

Codons are triplets of nucleotide bases that encode amino acids. Each position of an mRNA codon can be occupied by one of four bases: uracil (U), cytosine (C), adenine (A), or guanine (G). U and C are pyrimidines characterized by single-ring structures, whereas A and G are purines with double-ring structures. The mutation rate between codons depends primarily on the type of mutation and the specific position within the codon where the mutation occurs.

\begin{enumerate}
    \item Types of Mutation
\end{enumerate}

\begin{itemize}
    \item Transition mutation: A mutation that occurs between bases of the same structural type (purines or pyrimidines). Examples include adenine (A) changing to guanine (G), or cytosine (C) changing to uracil (U).
    \item Transversion mutation: A mutation that occurs between different structural types, from a purine to a pyrimidine, or vice versa. Examples include adenine (A) changing to cytosine (C) or uracil (U), and guanine (G) changing to cytosine (C) or uracil (U).
\end{itemize}

\begin{enumerate}
    \setcounter{enumi}{1}
    \item Codon Position
\end{enumerate}

\begin{itemize}
    \item First position: The first base of a codon plays an important role in determining the amino acid specified. Only leucine (Leu), arginine (Arg), and serine (Ser) have synonymous codons that differ at the first base.
    \item Second position: The second base of a codon is also crucial in determining the amino acid specified. Only serine (Ser) has synonymous codons that differ at the second base.
    \item Third position: Most of the redundancy in the SGC is attributable to the third base of a codon. This position is highly robust against transition mutations. Such robustness is demonstrated by the fact that transition mutations occurring at the third position are all synonymous, except for AUA (isoleucine) changing to AUG (methionine, start codon), and UGA (stop codon) changing to UGG (tryptophan). In contrast, transversion mutations at the third position, although more robust than mutations at the first or second positions, are considerably more likely to cause amino acid changes than transition mutations.
\end{itemize}

Denote the ratio of transition (ti) to transversion (tv) mutations in a nucleotide sequence as $\gamma \equiv \mathrm{ti}/\mathrm{tv}$. Under the assumption that all possible mutations among the four bases occur at equal rates, the probability of a transversion mutation would be expected to be twice that of a transition mutation, yielding a theoretical $\gamma$ value of 0.5. However, in reality, transition mutations occur more frequently than transversions, resulting in observed $\gamma$ values greater than 0.5 \cite{Duchene15}. For example, the $\gamma$ value for the entire nucleotide sequence of \textit{Drosophila} (fruit fly) is approximately 2 \cite{Begun07}, and for humans it is approximately 4 \cite{Hodgkinson10}. This indicates that, in humans, transition mutations occur about four times more frequently than transversion mutations across the entire nucleotide sequence.
\par
It is known that the GC content in the DNA of organisms ranges approximately from 20\% to 80\%. A plot of the overall genomic GC content on the x-axis versus the GC content at the three codon positions (P1, P2, and P3) on the y-axis reveals that the slopes are approximately in the ratio of $\mathrm{P1}:\mathrm{P2}:\mathrm{P3}=31:12:80$. This indicates that the third codon position is most sensitive to changes in the overall GC content. In line with the negative selection principle, which states that functionally less important regions evolve (change) more rapidly, it can be concluded that the third codon position, which is most affected by variations in the GC content, contributes the least to the specification of amino acids \cite{Saier19}. Consequently, from the perspective of error minimization, the mutation rate is inferred to be highest at the third codon position, followed by the first and second positions. This inference is further supported by the simulation model of Freeland and Hurst \cite{Freeland98, Freeland98_2}, which demonstrated that weighting mutations according to their positions increases the statistical improbability of the SGC.
\par
Hence the mutation rate between different codons can be calculated based on the type of mutation and the codon position. Assuming that mutations at each position occur independently, we can calculate the probability of a codon $XYZ$ mutating to $xyz$,

\begin{equation} \label{eq:1}
\mu_{(XYZ\rightarrow xyz)}=P(X\rightarrow x)\times P(Y\rightarrow y)\times P(Z\rightarrow z),
\end{equation}

where the values of the matrix $P$ are determined by the transition and transversion mutation weights at each position. The weights used in Freeland and Hurst's simulation model \cite{Freeland98} are given in Table~\ref{tab:table1}.

\begin{table}[h]
\caption{\label{tab:table1}%
The mutation weights used in prior study (Freeland \& Hurst, 1998).}
\begin{ruledtabular}
\renewcommand{\arraystretch}{1.25}
\begin{tabular}{cccc}
\textrm{}&
\textrm{1st position}&
\textrm{2nd position}&
\textrm{3rd position}\\
\colrule
For transitions & 1 & 0.5 & 1\\
For transversions & 0.5 & 0.1 & 1\\
\end{tabular}
\end{ruledtabular}
\end{table}

According to these values, the mutation rates for the $i$-th codon position, which is calculated as $w_i^{\mathrm{ti}}+2\times w_i ^{\mathrm{tv}}$, exhibit a ratio of $\mathrm{P1}:\mathrm{P2}:\mathrm{P3}=2:0.7:3$. Compared to the slopes of the GC content observed in the aforementioned plot, the ratio between P1 and P2 is similar, whereas the value for P3 is relatively lower. Although the slope ratios from the plot and the mutation rate ratios based on given weights are not necessarily identical, the comparison between the two datasets suggests the possibility that the mutation rate at the third position may have been underestimated. Notably, wobble pairing in codon-anticodon interactions is limited to the third position (wobble position) of the codon, further supporting this observation. Calculating the ratio of transition to transversion mutations, $\gamma$, using the values in Table~\ref{tab:table1}, yields a value of approximately 0.78, which is significantly lower than the $\gamma$ values reported for both humans and fruit flies. Therefore, we modify the weight for the third position transition, denoted hereafter as the “third transition” weight and abbreviated as $w_{\mathrm{tt}}\equiv w_3^{\mathrm{ti}}$, in our model to align with the existing data (Table~\ref{tab:table2}).

\begin{table}[h]
\caption{\label{tab:table2}%
The mutation weights used in this study.}
\begin{ruledtabular}
\renewcommand{\arraystretch}{1.25}
\begin{tabular}{cccc}
\textrm{}&
\textrm{1st position}&
\textrm{2nd position}&
\textrm{3rd position}\\
\colrule
For transitions & 1 & 0.5 & $w_{\mathrm{tt}}$\\
For transversions & 0.5 & 0.1 & 1\\
\end{tabular}
\end{ruledtabular}
\end{table}

According to the new weights in Table~\ref{tab:table2}, we obtain $w_{\mathrm{tt}}=4.9$ when $\gamma=2$ and $w_{\mathrm{tt}}=11.3$ when $\gamma=4$. Since the value of $\gamma$ is not universally consistent across species sharing the same genetic code (SGC) and even varies within the coding regions of genes in a single organism, it cannot be precisely determined for $w_{\mathrm{tt}}$. However, as indicated by the comparison of the SGC's robustness against different types of mutations, it is reasonable to expect that the third position would have a larger weight for transitions than for transversions. Consequently, statistical analyses of the genetic code will be conducted for $w_{\mathrm{tt}} \ge 1$.

\subsection{Codon translation errors}
\label{subsec:error_load}

Calculating translation errors in the genetic code first requires measuring the distances between amino acids. Several previous studies on error load minimization of the genetic code \cite{Freeland98, Freeland98_2, Goldman93, DiGiulio89, Alff-Steinbeger69} define amino acid distance based on differences in polar requirements. The polar requirement, $\phi$, is a value derived from the chromatographic behavior of amino acids in dimethylpyridine (DMP) solution \cite{Woese66, Mathew08},

\begin{equation} \label{eq:2}
\phi=-\left[\frac{d\left(\ln{R_m}\right)}{d\left(\ln{\chi_w}\right)}\right],~~R_m=\frac{1-R_f}{R_f}=\frac{\nu_{\mathrm{solvent}}-\nu_{\mathrm{sample}}}{\nu_{\mathrm{sample}}},
\end{equation}
\\
where $R_f$ is the retardation factor, $\nu_{\mathrm{sample}}$ is the distance traveled by the substance (solute), and $\nu_{\mathrm{solvent}}$ is the distance traveled by the solvent front. In other words, the polar requirement is defined as the negative of the slope of a log-log plot with $\chi_w$ (the mole fraction of water) on the x-axis and $R_m$ on the y-axis. The fact that this slope takes on distinct values for different amino acids can be explained by the interactions between organic bases and amino acids. During the dissolution of an amino acid in the solvent, interactions arise between the two, with both polar and nonpolar forces contributing to the overall solvation process. If the amino acid and the solvent interact relatively strongly in a nonpolar manner, the need for polar interactions diminishes. Therefore, certain amino acids with significantly large aliphatic R groups, such as Leucine, are expected to have a shallower slope, i.e., lower $\phi$ compared to other amino acids such as Alanine \cite{Woese66}.

\begin{figure}[t]
\includegraphics[width=0.45\textwidth]{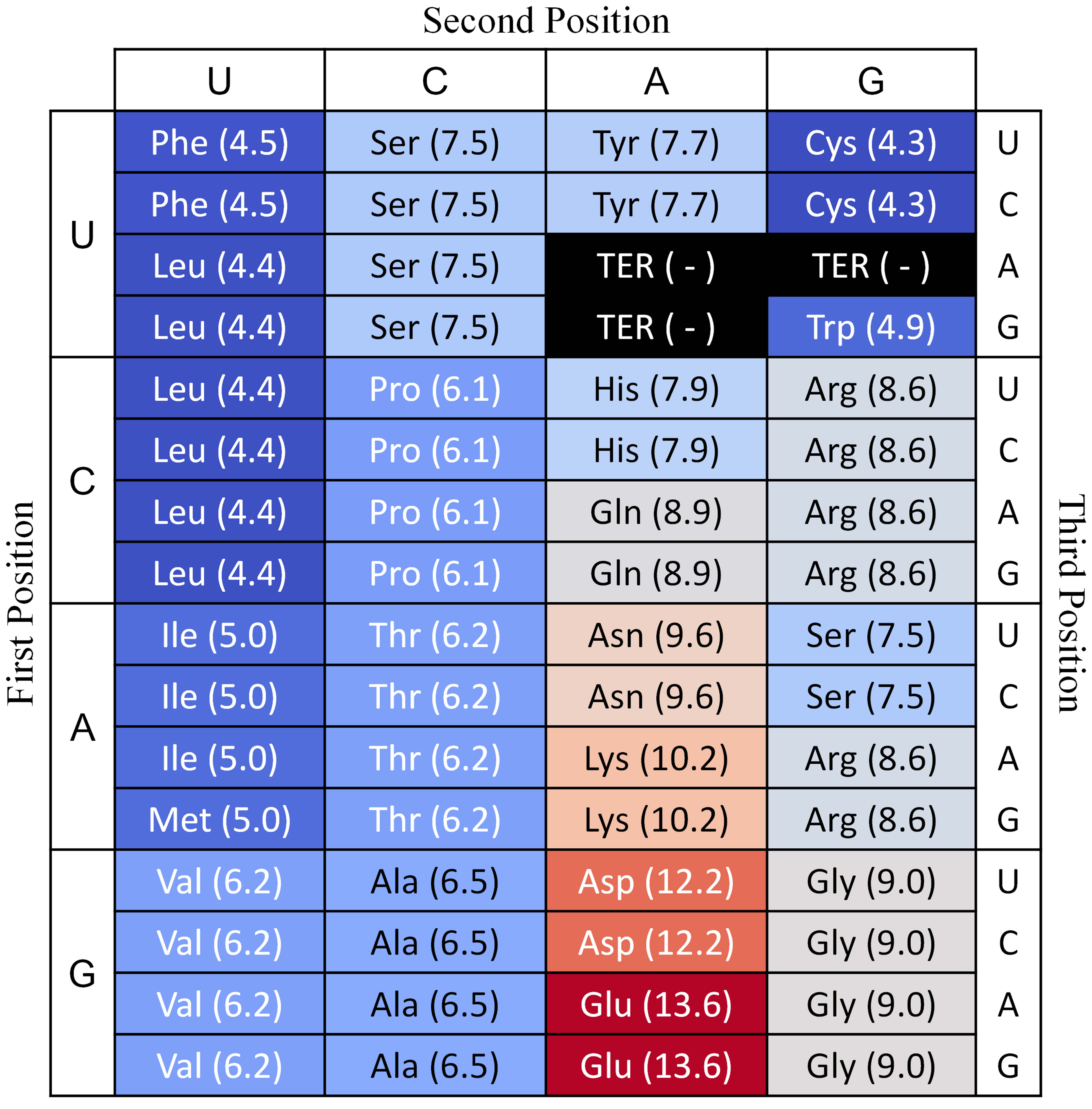}
\caption{\label{fig:FIG1} Codon table colored according to the polar requirement of the amino acids encoded by each codon, with blue indicating low values and red indicating high values. Neighboring codons tend to specify amino acids with similar polar requirement values, illustrating the nonrandom organization of the genetic code.}
\end{figure}

\begin{figure*}[t]
\includegraphics[width=\textwidth]{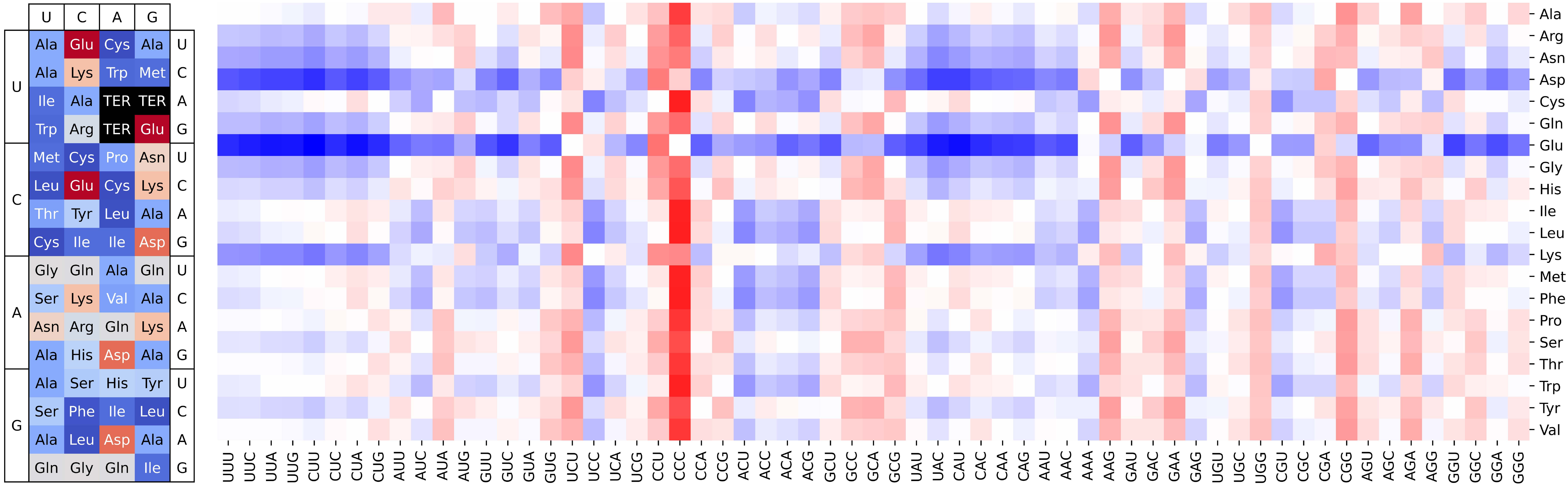}
\caption{\label{fig:FIG2}Randomly generated genetic code in Fig~\ref{fig:FIG1} format (left) and the heatmap representation of the error load of the corresponding local variation (right). In the heatmap, each column represents a codon subjected to the variation, while each row corresponds to an amino acid newly assigned due to deviations from the original codon assignment. Red cells indicate reduced error load (improvement), whereas blue cells indicate increased error load (deterioration). For instance, the amino acid Glu (glutamic acid), encoded by the codon CCC, exhibits a polar requirement significantly different from those of amino acids encoded by neighboring codons, especially for third position transitions; thus, most neighboring codes involving CCC show improvements. With 61 sense codons and 19 possible amino acid substitutions per codon, the total number of possible neighboring codes is $61\times19=1159$.}
\end{figure*}

The pattern of the SGC shown in Fig.~\ref{fig:FIG1} suggests that codon assignments are highly correlated with the values of polar requirements. Based on this characteristic, the error load of the genetic code can be defined in terms of differences in polar requirement, denoted as $\vert\Delta\phi\vert$. Previous studies assessing the error minimization of the genetic code have typically used absolute error \cite{Alff-Steinbeger69} or the squared error \cite{Freeland98, Goldman93, DiGiulio89} of polar requirement differences \cite{Ardell98}. This concept can be generalized such that for a positive real number $n$ \cite{Freeland00}, the error load takes the form: 

\begin{equation} \label{eq:3}
E=\sum_{i, j}\mu_{ij}\vert\phi_i-\phi_j\vert^n.
\end{equation}

Here, $\mu_{ij}$ represents the mutation rate between two codons $i$ and $j$, determined by the six mutation weights mentioned in Section~\ref{subsec:weights}. For instance, consider calculating the error load between the codons UUC and AUU. Assuming the probability of a point mutation occurring at a single codon position is $c$, the mutations between these two codons involve a transversion at the first position and a transition at the third position. Therefore, substituting the relevant weights from Table~\ref{tab:table1}, $\mu_{ij}$ becomes $w_1^{\mathrm{tv}} c \times (1-w_2^{\mathrm{ti}} c-2w_2^{\mathrm{tv}}c) \times w_3^{\mathrm{ti}} c= 0.5c^2-0.35c^3$. Even in environments where mutation weights differ, the probability $c$ remains constant. Thus, by dividing each weight by a normalization constant $k = (1 + 0.5 + 1) + 2 \times (0.5 + 0.1 + 1) = 5.7$, it becomes possible to perform appropriate numerical comparisons when varying the value of $w_{\mathrm{tt}}$. In the studies by Freeland and Hurst \cite{Freeland98, Freeland98_2}, the statistical improbability of the SGC was estimated through simulations that generated random genetic codes while preserving the synonymous codon blocks of the SGC. These simulations evaluated error load using the squared error, i.e., with $n=2$. Notably, in simulations employing the weights presented in Table~\ref{tab:table1}, the frequency of codes outperforming the SGC was found to be only about one in a million. In other words, the extremely low frequency of more efficient codes---even when the simulations were confined to a local neighborhood preserving synonymous codon blocks rather than exploring the entire coding space---suggests that the SGC is a highly efficient code in terms of error load minimization.
\par
Nevertheless, there is no necessity to define the amino acid distance using the squared error of polar requirement. Indeed, the statistical improbability of the SGC remains evident even when amino acid distances are defined using absolute error or arbitrary values of $n$ \cite{Freeland00}. Moreover, amino acid distance is not restricted to definitions based solely on polar requirement; alternative measures include the Grantham matrix \cite{Grantham74}, the Miyata matrix \cite{Miyata79}, the EMPAR matrix \cite{Rao87}, the HDSM matrix \cite{Prlic00}, the EX matrix \cite{Yampolsky05}, and the PAM matrix \cite{Ardell98, Dayhoff78}, which are computed from various biochemical properties of amino acids (e.g., molecular volume, composition, hydrophilicity) \cite{Massey16}. However, sequence-based matrices, such as PAM, inherently depend on SGC, potentially introducing biases. Therefore, physicochemically defined metrics that are independent of the SGC, such as the polar requirement, may provide more appropriate and unbiased measures for modeling the evolution of the genetic code \cite{Novozhilov07}.
\par
In this study, we focus on analyzing genetic codes differing from the SGC by a single codon reassignment, each referred to as a \textit{neighboring code}. Each neighboring code thus differs from the SGC in exactly one codon-to-amino acid assignment. We use the term \textit{local variation} more generally to describe the collective set of these single-site modifications, as well as to refer to analyses exploring the structural consequences of such single-step deviations. For generality, we define the error load for arbitrary $n$ using polar requirement as the amino acid distance metric. Then, we conduct the local variation analysis (an example is shown in Fig.~\ref{fig:FIG2}) to examine whether improvements in the genetic code efficiency exist through deviations from the SGC. Specifically, we consider all possible amino acid substitutions for each individual codon to identify the most appropriate value of $n$.
\par
As mentioned in Section~\ref{subsec:weights}, the value of $\gamma$ is typically greater than 1.5. When $w_{\mathrm{tt}}$ becomes very large, only genetic codes robust against transition errors at the third codon position can become local optima. Therefore, for any genetic code to evolve toward optimality, synonymous codons connected by a transition mutation at the third position must pair up and undergo coordinated changes. In other words, to effectively identify local optima through step-by-step amino acid substitutions, the local optimum must be achievable at a reasonably small value of $w_{\mathrm{tt}}$, assuming a “natural” error metric is employed. Consequently, we will examine the suitability of the genetic code using squared error, absolute error, and the generalized $n$-power error metric, by exploring local variation around the SGC and gradually increasing the value of $w_{\mathrm{tt}}$ from 1, to determine the minimal point at which the code becomes locally optimal.

\subsection{Amino acid compositional alignment}
\label{subsec:amino_acid_KLD}

The error load of the genetic code is determined by the mutation rates ($\mu_{ij}$) between codons and the distances between their encoded amino acids, which inherently depend on the graph structure of the code itself \cite{Tlusty08, Tlusty10, Choi17}. If code optimality were defined solely by minimizing error load, the global optimum would trivially occur when all codons specify a single amino acid ($\Delta\phi = 0$, $E = 0$), regardless of model parameters. However, for organismal survival, all 20 amino acids must be represented. Furthermore, frequently used amino acids are plausibly encoded by multiple codons, which helps reduce the impact of translational errors. Therefore, the relative abundances of the 20 amino acids should be incorporated when evaluating code optimality.

Incorporating this logic into our framework provides insight into the nonuniform codon assignments observed in nature.
While previous studies considered the diversity of the amino acid vocabulary using simplified or qualitative measures of diversity~\cite{Tlusty08,Tlusty2008c}, here we explicitly account for the nuanced distribution of amino acids observed in present-day organisms. 
This consideration motivates the introduction of an amino acid Kullback-Leibler divergence (KLD) term to quantify the evolutionary pressures exerted by varying amino acid demands.

Hence, recognizing that each amino acid is utilized in differing quantities during protein synthesis, we define the amino acid KLD term ($D_{\mathrm{KL}}$) as follows:

\begin{equation} \label{eq:4}
D=D_{\mathrm{KL}}\left(f_\alpha\parallel p_\alpha\right)=\sum_{\alpha\in\mathrm{AA}}f_\alpha\ln{\frac{f_\alpha}{p_\alpha}}.
\end{equation}
Here, $f_\alpha$ denotes the frequency of amino acid $\alpha$ in the proteins of an organism, where $\alpha\in\mathrm{AA}$ and $\mathrm{AA}$ denotes the set of the 20 canonical amino acids. Although various datasets have been previously reported \cite{Dayhoff69, Brooks02, Jungck78, Jukes75, Dayhoff78_2}, in this study, we calculate amino acid frequencies from the coding DNA sequences (CDS) of genome data to investigate organisms with different genetic codes. $p_\alpha$ denotes the proportion of codons that specify amino acid $\alpha$ in a given code. Consequently, minimizing the amino acid KLD requires that the genetic code must reflect the actual proportions of amino acids. However, while KLD quantifies the divergence between two probability distributions, it is asymmetric and thus does not strictly satisfy the mathematical definition of a distance. Therefore, when defining amino acid KLD, a choice must be made between $D_\text{KL}\left(f_\alpha\parallel p_\alpha\right)$ and $D_\text{KL}\left(p_\alpha\parallel f_\alpha\right)$. Given that $f_\alpha$ represents the true distribution and $p_\alpha$ represents the estimated distribution, it is logical to adopt the forward KLD definition, $D_\text{KL}\left(f_\alpha\parallel p_\alpha\right)$, as all regions with $f_\alpha>0$ must also be covered by $p_\alpha$. KLD can be expressed as the difference between the cross entropy $H_p(f)=-\sum{f_\alpha\ln{p_\alpha}}$ and the Shannon entropy of $f_\alpha$, $H=-\sum{f_\alpha\ln{f_\alpha}}$. Since $H$ is constant, cross entropy alone could be used instead of KLD. Nevertheless, for numerical comparisons with the error load, it is beneficial to have a measure that attains a minimum value of zero, motivating the choice of KLD to quantify the similarity between the two amino acid distributions. Ultimately, the loss function of the genetic code is defined as a linear combination of error load $E$ and amino acid KLD $D$,

\begin{eqnarray} \label{eq:5}
L&=&E+\eta D\nonumber\\&=&\sum_{i, j}\mu_{ij}\vert\phi_i-\phi_j\vert^{n}+\eta\sum_{\alpha\in\mathrm{AA}}f_{\alpha}\ln\frac{f_\alpha}{p_\alpha},
\end{eqnarray}
where $\eta$ serves as a balancing parameter that adjusts the relative contributions of $E$ and $D$. Specifically, if $E$ is interpreted as the cumulative error within the code and $D$ as the external selective pressure to satisfy amino acid demand beyond the code, then tuning $\eta$ allows us to assess which factor---$E$ or $D$---exerts a more significant influence on the evolution of the genetic code at any given point in time.

\section{Results}
\label{sec:results}

\subsection{The genetic code achieves optimality through local variation}
\label{subsec:local_variation}

In this experiment, the probability of a point mutation was set to $c=10^{-10}$ \cite{Balin10}. The six types of mutation weights were normalized to preserve the overall mutation probability $c$, even if $w_{\mathrm{tt}}$ varies. The mutation rate $\mu_{ij}$ between two codons $i$ and $j$ is zero when $i = j$, and, since $c$ is a very small value, the contribution of codon mutations involving changes at two or more positions to the error load is negligible compared to point mutations. Accordingly, if we consider only point mutations, the threshold value of the third transition weight, $w_{\mathrm{tt}}^\ast$, at which the error load of the SGC becomes a local optimum, can be approximated. In this study, amino acid distances are defined using both absolute error and squared error (see Appendix for more details).

\begin{itemize}[leftmargin=*]
\item Absolute error
\begin{equation} \label{eq:6}
w_{\mathrm{tt}}^\ast\approx\max\left(\sum_{i\in\aleph}w_i\right)
\end{equation}
\end{itemize}

\begin{itemize}[leftmargin=*]
\item Squared error
\begin{equation} \label{eq:7}
w_{\mathrm{tt}}^\ast\approx\max\left(\frac{2}{\delta}\sum_{i\in\aleph}w_i\tilde{\phi}_i-\sum_{i\in\aleph}w_i\right)
\end{equation}
\end{itemize}

\begin{figure}[t]
\includegraphics[width=0.48\textwidth]{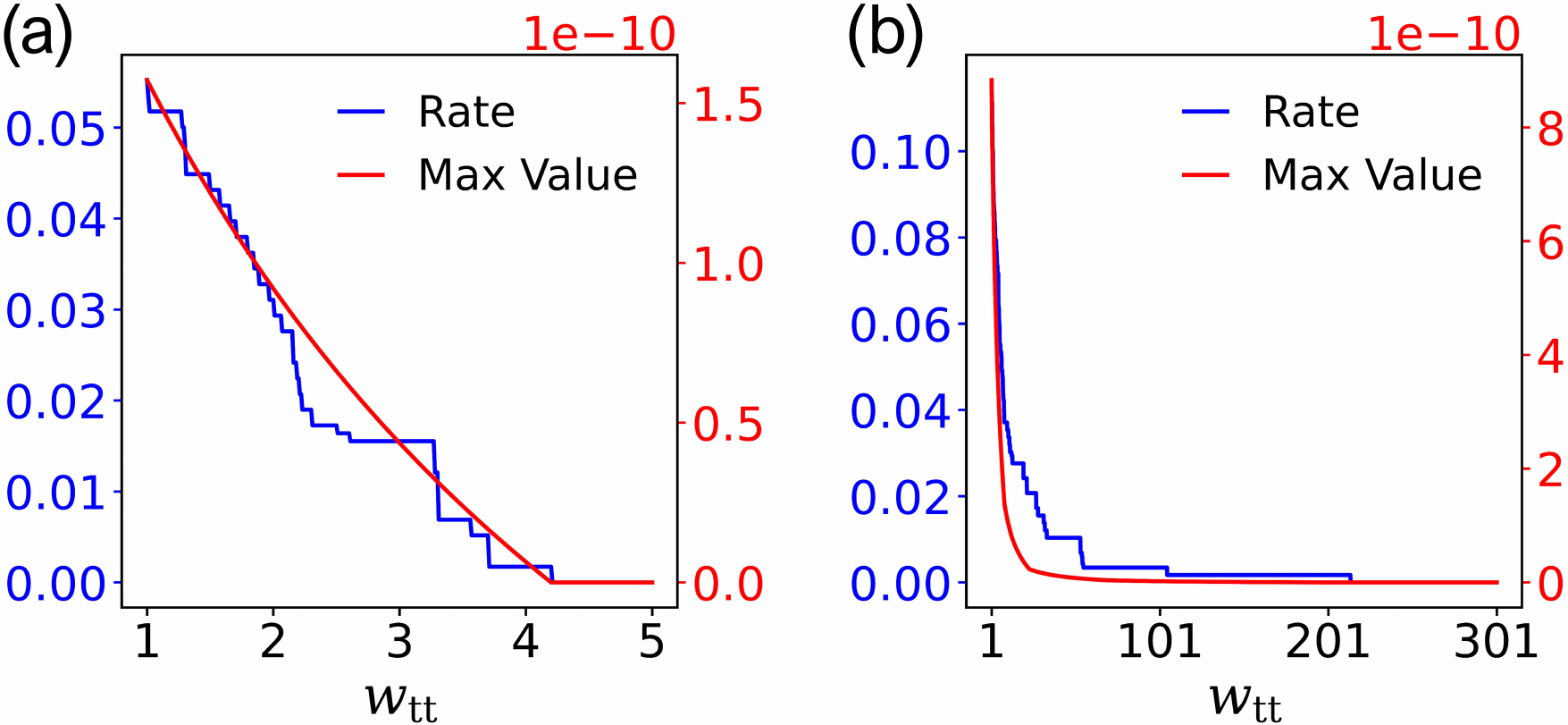}
\caption{\label{fig:FIG3} The error load of local variation around the SGC as the third transition weight ($w_{\mathrm{tt}}$) increases. (a) Amino acid distance defined as absolute error. (b) Amino acid distance defined as squared error. The blue line represents the improvement rate, defined as the fraction of neighboring codes exhibiting a lower error load compared to the SGC. The red line indicates the maximum improvement in error load achieved among these codes for each value of $w_{\mathrm{tt}}$.}
\end{figure}

\begin{itemize}[leftmargin=*]
\item $n$-power error
\begin{equation} \label{eq:8}
w_{\mathrm{tt}}^\ast\approx\max\left\{\sum_{i\in\aleph}w_i\sum_{k=1}^{n}(-1)^{k-1}\binom{n}{k}\bigg\vert\frac{\tilde{\phi}_i}{\delta}\bigg\vert^{n-k}\right\}
\end{equation}
\end{itemize}
Here, $\tilde{\phi}_i\equiv\phi_i-\phi_\alpha$, where $\phi_i$ denotes the polar requirement of the amino acid specified by codon $i$, and $\phi_\alpha$ is that of the amino acid originally specified by codon $\alpha$. The codon $\alpha$ refers to the codon whose assigned amino acid changes as a result of local variation. The set $\aleph$ represents the codons that are neighbors of $\alpha$, i.e., codons differing from $\alpha$ by a single base substitution, excluding those involved in a third position transition relationship and stop codons. $\delta$ denotes the difference in polar requirement between the original amino acid specified by codon $\alpha$ and the substituted amino acid. Accordingly, calculations show that, in the case of absolute error, $\alpha$ is either UGU or UGC, which both specify cysteine (Cys), yielding $w_{\mathrm{tt}}^\ast \approx 4.2$. In the case of squared error, $\alpha$ is either GAA or GAG, which both specify glutamic acid (Glu), yielding $w_{\mathrm{tt}}^\ast \approx 214.1$. These values precisely match the numerical experimental results presented in Fig.~\ref{fig:FIG3}.

The significantly larger value of $w_{\mathrm{tt}}^\ast$ in the case of squared error, compared to absolute error, can be explained by comparing Eq.~\eqref{eq:6} and Eq.~\eqref{eq:7}. Equation~\eqref{eq:7}, which describes $w_{\mathrm{tt}}^\ast$ under the squared error metric, includes a term inversely proportional to $\delta$. To maximize the expression inside the brackets, $\delta$ must be minimized; in this context, its value is 0.1. Since most of the $\tilde{\phi}_i$ values included in the summation are greater than 1, the term $2\delta^{-1}\sum_{i\in\aleph}w_i\tilde{\phi}_i$ can be estimated to be more than 20 times larger than the term $\sum_{i\in\aleph}w_i$. Therefore, the value of $w_{\mathrm{tt}}^\ast$ under squared error is concluded to be tens of times larger than that under absolute error. This property holds not only for the comparison between squared error and absolute error but also for the generalized $n$-power error, where the amino acid distance is defined as $\vert\Delta\phi\vert^n$. The value of $w_{\mathrm{tt}}^\ast$ in the generalized $n$-power error can be calculated using the binomial theorem, as shown in Eq.~\eqref{eq:8}. As a result, it includes a term proportional to $\vert\delta\vert^{n-1}$, which leads to a much larger $w_{\mathrm{tt}}^\ast$ value for $n>1$ compared to the case of absolute error.

\begin{figure}[t]
\includegraphics[width=0.48\textwidth]{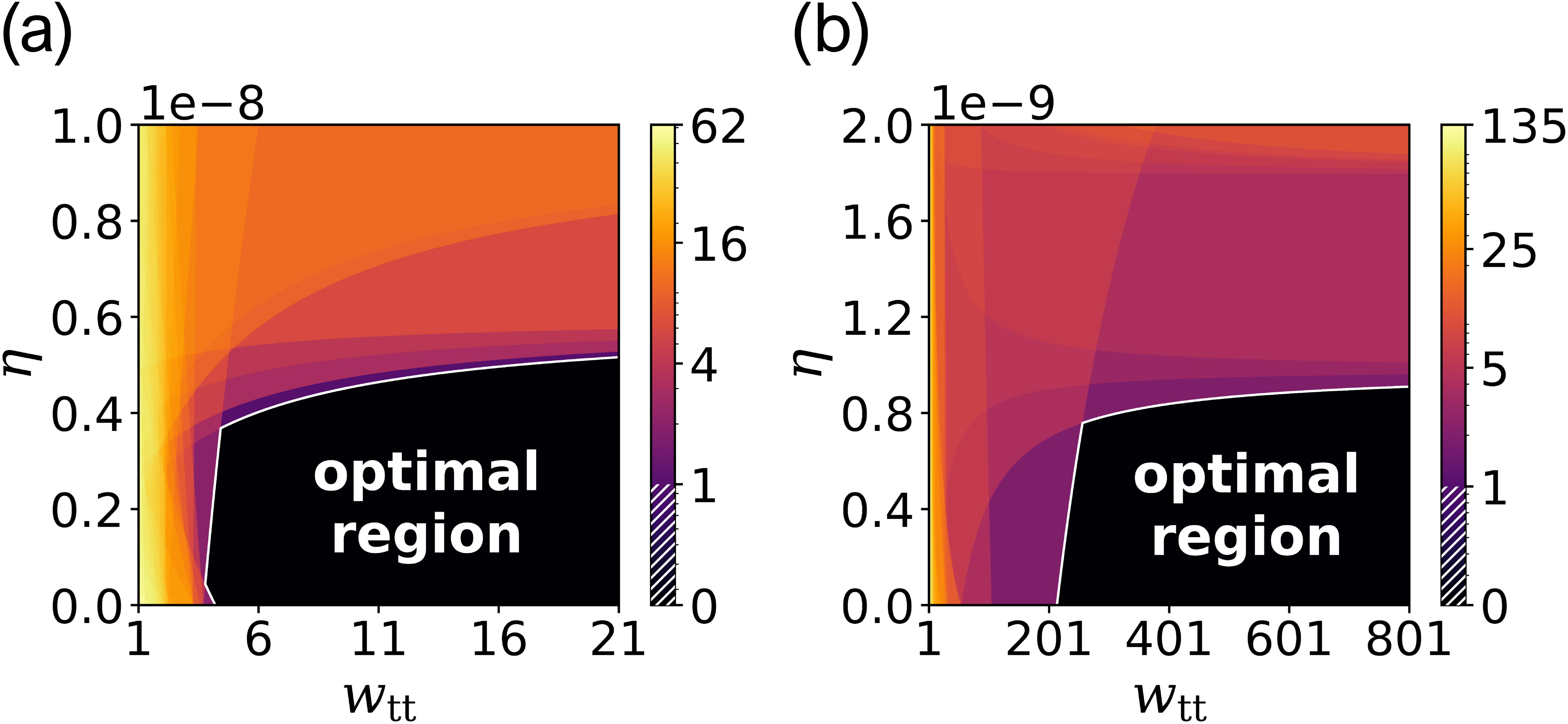}
\caption{\label{fig:FIG4} The loss landscape of local variation of the SGC as a function of changes in the third transition weight, $w_{\mathrm{tt}}$, and the balancing parameter, $\eta$. (a) Amino acid distance defined as absolute error. (b) Amino acid distance defined as squared error. The color bar represents the number of codes with a lower loss value than the SGC, presented on a logarithmic scale, out of the 1159 possible neighboring codes described in Fig~\ref{fig:FIG2}. Note that the interval $(0, 1)$ is omitted, and a value of 0 corresponds to the points where the SGC becomes a local optimum.}
\end{figure}

The experiment examining the error load of local variation can be extended in a similar manner to investigate the code loss of local variation, incorporating an additional amino acid KLD term. As shown in Fig.~\ref{fig:FIG4}, a heatmap can be generated over continuous ranges of $\eta$ and $w_{\mathrm{tt}}$, representing the ratio of the number of improved neighboring codes to the total number of possible neighboring codes compared to the SGC. In Fig.~\hyperref[fig:FIG4]{4(a)} (absolute error) and Fig.~\hyperref[fig:FIG4]{4(b)} (squared error), the boundary of the region where the SGC is locally optimal exhibits two generic features: a steep segment in the vicinity of $w_{\mathrm{tt}}\approx w_{\mathrm{tt}}^\ast$ and a gently sloped, nearly horizontal segment confined to a narrow band in 
$\eta$. The steep boundary is consistent with the observation that, for both metrics, the fraction of neighboring codes with lower loss than the SGC rises sharply as $w_{\mathrm{tt}}\rightarrow 1$. This behavior likely reflects a relaxation of the constraint that enforces robustness to third‐position transition mutations. By contrast, the gently sloped boundary is associated with the tendency to drift away from the local optimum as $\eta$ increases, and arises because the weights are normalized to keep the total point mutation probability fixed at $c$; for sufficiently large $w_{\mathrm{tt}}$, the error load term becomes dominated by third position transitions, rendering the boundary nearly horizontal in $w_{\mathrm{tt}}$.
\par
In conclusion, the key implications derived from the results of Fig.~\ref{fig:FIG4} are as follows:
\par
First, applying the absolute error with $n=1$ to the amino acid distance based on polar requirement $\vert\Delta\phi\vert^n$ ensures the optimality of the SGC under the most rational conditions. While the previous analysis of $w_{\mathrm{tt}}^\ast$ considered only the error load as a component of code loss, it has now been confirmed---through the local variation heatmap calculated under the complete definition of code loss, which includes the amino acid KLD term---that absolute error guarantees the optimality of the SGC at a much smaller $w_{\mathrm{tt}}^\ast$ value compared to squared error.
\par
Second, for the SGC to be a “strict” local optimum, the balancing parameter $\eta$ must be low, even after accounting for the difference in scale between the two terms. This indicates that the contribution of error load is greater than that of amino acid KLD in optimizing the genetic code. In particular, examining the regions where the SGC becomes a local optimum in the heatmaps of Fig.~\ref{fig:FIG4} suggests that the selective pressure to minimize amino acid KLD operates independently of the SGC's optimality. This observation is also evident in the experimental results presented in Section~\ref{subsec:simulated_annealing} and Section~\ref{subsec:distribution}.
\par
However, the preceding analysis is limited to local variation around the SGC and is therefore not applicable in the broader context of code evolution, where parameters influencing code loss change over time and the genetic code itself evolves. To gain insights into the evolution of the genetic code, it is essential to understand the detailed characteristics of the global loss landscape and how it may shift over evolutionary timescales. Accordingly, in our subsequent experiments, we aim to explore the properties of optimal codes within the loss landscape and to investigate how these codes are influenced by changing environmental factors during the evolutionary process.

\subsection{Simulated annealing balances translation error and amino acid similarity}
\label{subsec:simulated_annealing}

To go beyond local variation around the SGC and fully address the loss landscape of genetic codes, it is crucial to characterize the properties of local optima. From an evolutionary perspective, natural selection typically favors directions that decrease loss, eventually driving the genetic code toward a local optimum. Therefore, understanding the distribution of these local optima may offer insights into the evolutionary constraints shaping the current structure of the SGC. However, examining each of the numerous optima dictated by the specific topology is a challenging task. Thus, we first investigate the general statistical properties of code loss and subsequently apply these insights to explore the landscape.
\par
The optimization of the genetic code is a combinatorial problem in which the number of potential solutions increases exponentially. Due to the complex topology of the loss landscape, there exist multiple local optima that vary significantly in their proximity to the global optimum: some possess loss values comparable to the global optimum, while others represent substantially lower loss regions. Considering these characteristics, we adopt simulated annealing to efficiently explore the loss landscape and identify solutions closer to the global optimum. By probabilistically accepting solutions that temporarily worsen the objective function, particularly at high initial temperatures, simulated annealing reduces the likelihood of becoming confined to poorly performing local optima. As the temperature gradually decreases according to a predefined schedule, the algorithm progressively converges toward stable, high-quality solutions. This stochastic search mechanism is therefore expected to outperform traditional deterministic methods in evaluating the error-minimizing properties of the genetic code.

\algnewcommand{\Inputs}[1]{%
  \State \textbf{Inputs:}
  \Statex \hspace*{\algorithmicindent}\parbox[t]{.8\linewidth}{\raggedright #1}
}
\algnewcommand{\Initialize}[1]{%
  \State \textbf{Initialize:}
  \Statex \hspace*{\algorithmicindent}\parbox[t]{.8\linewidth}{\raggedright #1}
}

\newfloat{algorithm}{h}{lop}
\begin{algorithm}
\caption{Simulated annealing}\label{alg:cap}
\begin{algorithmic}[1]
\Inputs{Standard deviation of random code data: $\sigma$ \\ 
        Number of iterations: $n$ \\ 
        Initial temperature: $T_i = 10^{-4}\sigma$ \\ 
        Stop temperature: $T_s$ \\ 
        Cooling rate: $\beta \gets (T_s/T_i)^{1/n}$}
\Initialize{$T\gets T_i$ \\ 
            Generate a random initial code: $C_R$}
\While{$T > T_s$}
    \State Generate a random neighboring code of $C_R$: $C_N$
    \State Calculate loss: $L(C_R)$, $L(C_N)$
    \If{$L(C_R)\geq(C_N)$}
        \State $p\gets1$
    \Else
        \State $p\gets\exp\left[T^{-1}\bigl(L(C_R)-L(C_N)\bigr)\right]
$
    \EndIf
    \State With probability $p$, set $C_R\gets C_N$
    \State $T\gets\beta\cdot T$
\EndWhile
\end{algorithmic}
\end{algorithm}

The optimization algorithm employed in this experiment follows the standard framework of simulated annealing, with several hyperparameters adjusted to ensure consistent optimization conditions across different $\eta$ values. Namely, the initial temperature was set proportionally to the standard deviation of loss calculated from one million randomly generated codes. This approach ensures that the balance between exploration and exploitation remains consistent, even when the dispersion of the loss landscape varies according to the value of $\eta$. Additionally, the cooling rate was defined as the $n$-th root of the total number of iterations, thereby maintaining the total number of iterations constant at $n$.
\par
In the optimization process, the value of $\eta$ was determined based on z-scores calculated independently for the error load ($E$) and amino acid KLD ($D$) of the SGC. The mean and standard deviation required for the calculation were obtained from the distributions of $E$ and $D$ derived from the ensemble of random codes (Fig.~\ref{fig:FIG6}). Specifically, under the assumption that the SGC represents an optimal code, the extent to which its $E$ and $D$ values deviate from their respective means was interpreted as reflecting the complexity of each landscape. Consequently, the ratio of the corresponding z-scores, defined as $\tilde{\eta}=z(D_\text{SGC})/z(E_\text{SGC})\approx0.56$, was adopted as an ideal estimate of $\eta$. Furthermore, to investigate how variations in $\eta$ influence the optimization process, additional candidate values for $\tilde{\eta}$ ($10^{-3}, 10^{-2}, 10^{-1}, 10^{0}, 10^{1}, 10^{2}, 10^{3}$) were considered in this study. By applying simulated annealing under identical conditions across these candidate values, we examined how different choices of $\eta$ influence both the resulting optimal codes and the corresponding optimization trajectories. The outcomes of these analyses provide insight into the role of $\eta$ in interpreting the evolutionary processes underlying the genetic code.
\par
Through analysis of the local variation presented in Fig.~\ref{fig:FIG4}, we observed that the value of $\eta$ corresponding to the condition under which the SGC resides at a local optimum is remarkably low. In particular, as $\eta$ increases from zero, the optimality of the SGC is maintained through an increase in $w_{\mathrm{tt}}^\ast$. This suggests that, at least for the specific local optimum occupied by the SGC, an increase in $\eta$---that is, selective pressure exclusively aimed at minimizing $D$---is insufficient to fully explain the current structure of the SGC. However, it is important to acknowledge that the loss function defined in this study serves merely as an approximate representation of the selective pressures acting during the evolutionary history of the genetic code. Consequently, the range of “appropriate” $\eta$ values derived from this function within the context of local variation should not be regarded as absolute. Rather than strictly determining whether a given code precisely corresponds to an optimal code in this framework, we employ a relative measure to evaluate how closely a code approximates such an optimal state.

\begin{figure*}[t]
\includegraphics[width=0.9\textwidth]{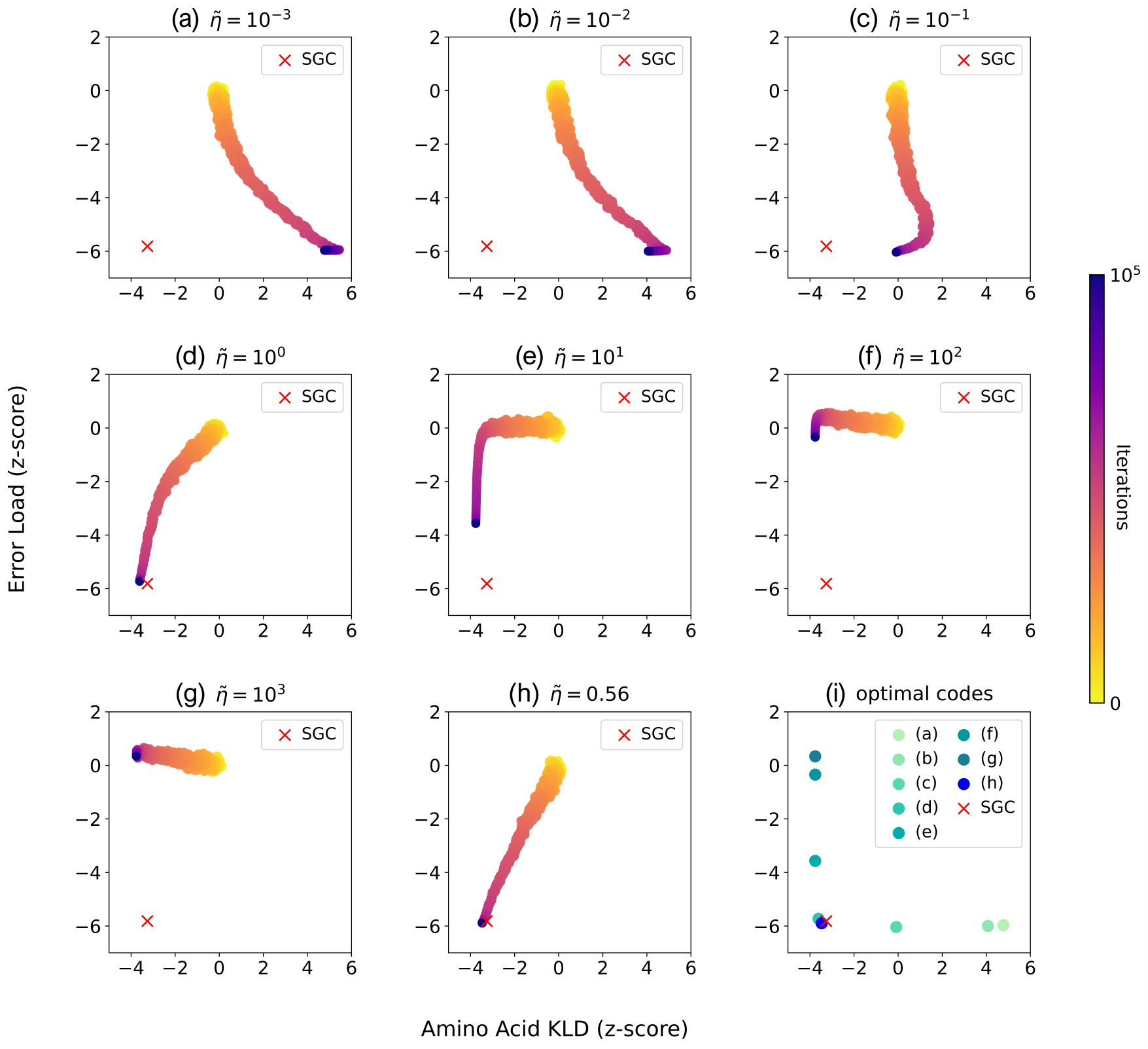}
\caption{\label{fig:FIG5} (a)–(h) Results of optimization via simulated annealing for eight distinct values of $\tilde{\eta}$. Here, we define $\tilde{\eta}\equiv(\sigma_D/\sigma_E)\eta\approx 4\times10^7\eta$, where $\sigma_D$ and $\sigma_E$ are the standard deviations of the amino acid KLD and the error load, respectively, obtained from the ensemble of random codes. Using $\tilde{\eta}$ ensures that when $\eta$=1, the optimization assigns equal relative importance to minimizing both the error load and amino acid KLD. All points represent standardized z-scores of the error load ($E$, vertical axis) and amino acid KLD ($D$, horizontal axis). Each optimization trajectory corresponds to the average trajectory obtained from 100 simulated annealing runs, with the initial state being a randomly generated code. As in previous local variation experiments, optimization was conducted on 61 sense codons (excluding the stop codons) and based on the natural frequency of \textit{Homo sapiens}. The optimization target (position of the SGC) is indicated by a red cross. (i) Final optimal codes obtained at the last iteration of simulated annealing for each of the eight $\tilde{\eta}$ values shown in panels (a)–(h). The result corresponding to the ideal value ($\tilde{\eta}=0.56$) is highlighted in blue.}
\end{figure*}

The experimental results presented in Fig.~\ref{fig:FIG5} demonstrate that, depending on the magnitude of $\eta$, there is a trade-off between the error load $E$ and the amino acid KLD $D$ of the optimal codes. Specifically, at high $\eta$ values (Figs.~\hyperref[fig:FIG5]{5(f)}, \hyperref[fig:FIG5]{5(g)}), $D$ is minimized very effectively, reaching values even smaller than $D_{\mathrm{SGC}}$, while the reduction in $E$ remains almost negligible. In contrast, at low $\eta$ values (Figs.~\hyperref[fig:FIG5]{5(a)}, \hyperref[fig:FIG5]{5(b)}), $E$ is substantially minimized, attaining values comparable to or even smaller than $E_{\mathrm{SGC}}$, whereas $D$ tends to increase significantly beyond its initial (random average) value prior to optimization. This phenomenon arises because, as described in Section~\ref{subsec:amino_acid_KLD}, when $\eta$ is sufficiently low for $D$ to be disregarded, $E$ is minimized by assigning most codons to a single amino acid. Conversely, this results in highly uneven code frequencies, leading to a substantially elevated value of $D$.

\subsection{Genetic codes of species occupy exceptionally rare and optimal regions}
\label{subsec:distribution}

In the preceding analysis using simulated annealing, we confirmed that the optimal codes converge to local optima close to the SGC when the balancing parameter is set to $\tilde{\eta}\approx0.56$. However, since some organisms employ genetic codes that differ slightly from the SGC, and the natural amino acid frequency $f_\alpha$ varies across species, this optimization process---based solely on the SGC, and specifically on data from \textit{Homo sapiens}---offers a limited view of genetic code evolution. To address this limitation, we perform random code simulations analyzing both error load and amino acid KLD. Our goal is to reexamine the structural features of various genetic code variants---including the SGC---and to assess how much the minimization of error load $E$ and amino acid KLD $D$ may have shaped their evolution. In this experiment, random codes are generated according to the following procedure.

\begin{enumerate}
  \item Let $m=64-n$ be the number of sense codons. Randomly permute these $m$ codons.
  \item Uniformly choose $19$ cut points among the $(m-1)=(63-n)$ gaps between adjacent codons, thereby partitioning the list into $20$ contiguous blocks (i.e., sampling an ordered composition of $m$ into $20$ positive parts; $\binom{63-n}{19}$ choices).
  \item Assign the $i$-th block to the $i$-th amino acid in a fixed order to obtain a code.
\end{enumerate}

\begin{figure}[t]
\includegraphics[width=0.48\textwidth]{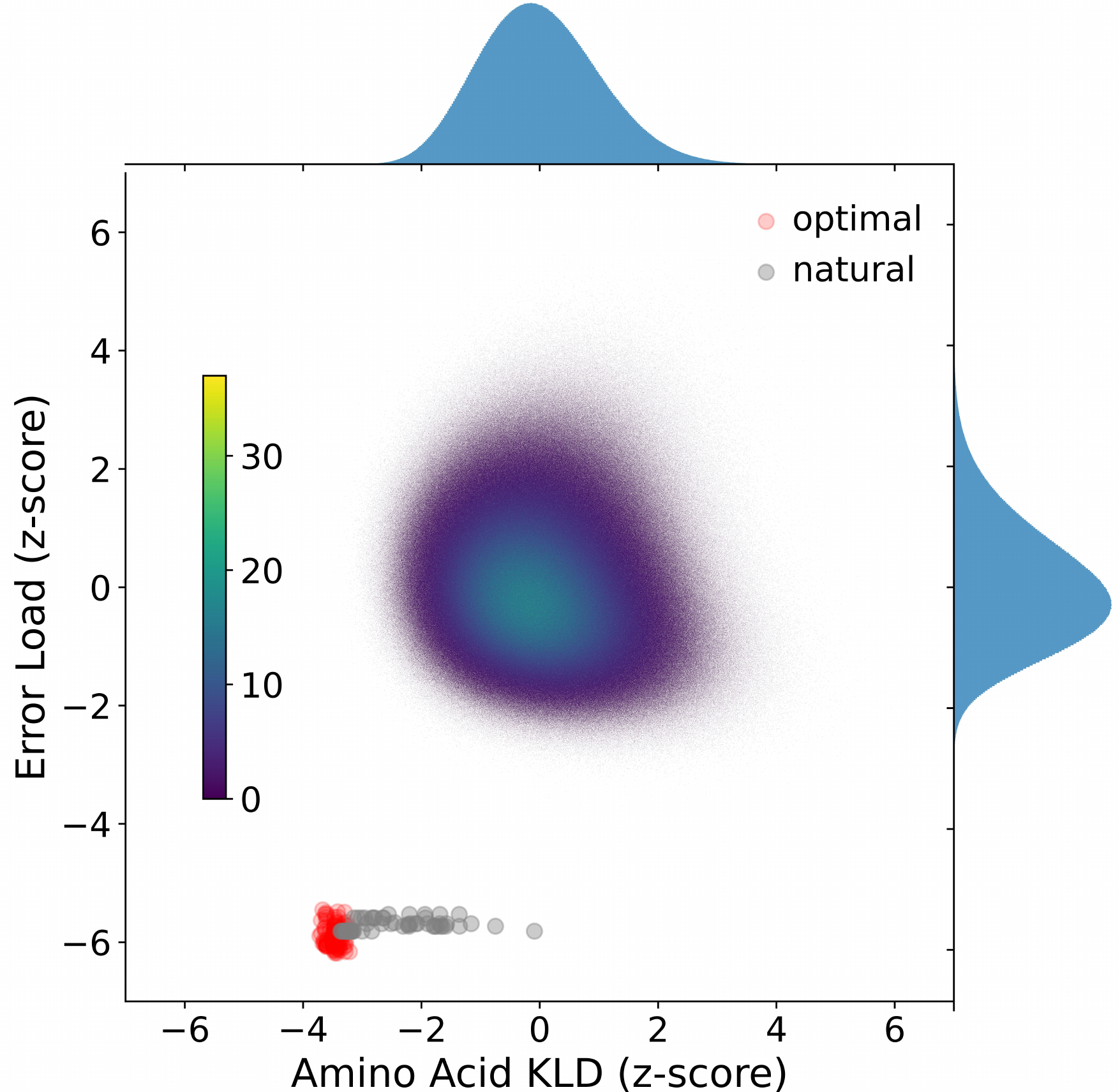}
\caption{\label{fig:FIG6} Distribution of error load versus amino acid KLD for genetic codes. Joint distribution showing $5.8\times10^7$ random codes (density heatmap), 58 natural codes (gray circles), and 100 optimal codes (red circles). Random codes were generated by randomly partitioning the $m$ sense codons into 20 amino acid groups, with $10^6$ codes sampled for each of the 58 species. Marginal histograms show the distribution of error load (right) and amino acid KLD (top). Both metrics are presented as z-scores normalized by the mean and standard deviation of the random code distribution. Natural codes cluster in the lower-left region, indicating simultaneously low error load and low amino acid frequency bias compared to random expectation.}
\end{figure}

Both $E$ and $D$ obtained from random codes exhibit positively skewed, approximately skew-normal distributions (Fig.~\ref{fig:FIG6}). The code distribution presented in Fig.~\ref{fig:FIG6} shows that there is no significant correlation between error load and amino acid KLD in randomly generated codes (Pearson's $r=-0.139$); in other words, optimization of one term is nearly independent of the optimization of the other. Under our randomization scheme, the code frequency vector is an ordered composition of $m$ codons into 20 amino acids ($p_\alpha=k_\alpha/m\ge 1/m$ with $\sum_\alpha k_\alpha=m$), which yields a block-size distribution heavily weighted toward small blocks. Consequently, terms with small $p_\alpha$ inflate $D=\sum_\alpha f_\alpha\ln(f_\alpha/p_\alpha)$, producing a hard lower bound at $0$ and a long right tail. For $E$ at $n=1$, i.e., $E=\sum_{i,j}\mu_{ij}\,|\phi_i-\phi_j|$, low values require globally coherent clustering of similar polar requirements across the codon-adjacency graph; such arrangements occupy a vanishingly small fraction of code space, so random assignments concentrate at moderate–high $E$ with an even longer right tail than $D$. These combinatorial and geometric constraints jointly explain the observed skewness, which is empirically larger for $E$ than for $D$ (approximately 2.1 vs. 1.4 in our data).

\begin{figure*}[t]
\includegraphics[width=\textwidth]{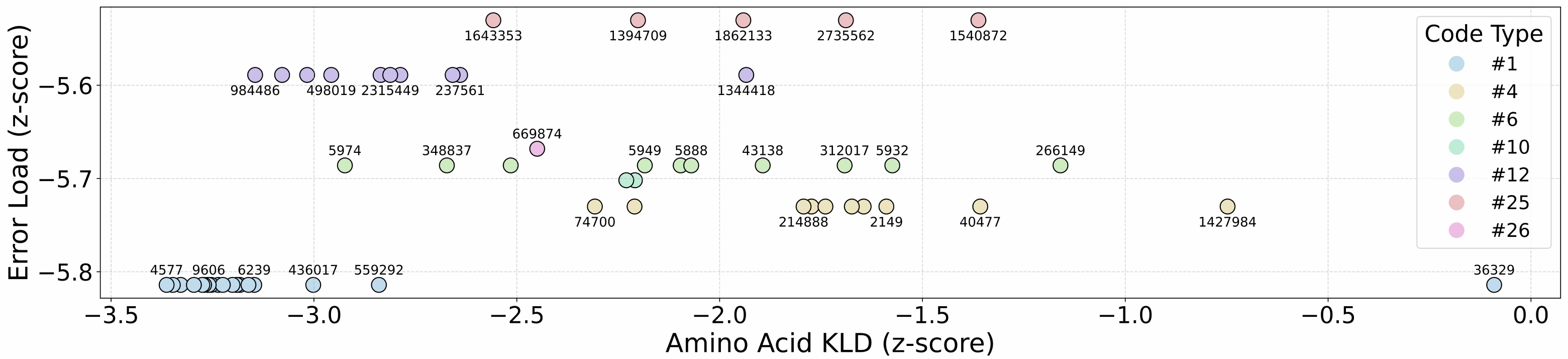}
\caption{\label{fig:FIG7} 
Scatter plot of standardized error load $E$ and amino acid KLD $D$ values for 58 species with variant genetic codes. This plot shows an enlarged view of the region in the code distribution space (Fig.~\ref{fig:FIG6}) where the natural code is located. Numbers adjacent to each data point indicate the corresponding NCBI taxonomy ID for each species (detailed in Table~\ref{tab:ncbi_data}). Different colors represent distinct genetic code types, with seven code variants analyzed including the SGC (type 1). The human genetic code (\textit{Homo sapiens}, ID: 9606) is located in the lower left region of the plot.}
\end{figure*}

Figure~\ref{fig:FIG7} displays standardized z-scores of the error load ($E$) and amino acid KLD ($D$) for 58 species. A first, code-level regularity is that $E$ is identical within each genetic code type ~\cite{Suzuki21} (constant horizontal bands), reflecting that $E$ depends only on the codon-amino acid mapping, not on species-specific amino acid usage. Among the surveyed codes, the SGC (type 1) attains the most negative (lowest) z-score for $E$ and thus the highest error robustness, while variant codes show progressively less negative $E$ (e.g., types 4, 6, 10, 12, 25, 26). This pattern indicates that the observed codon reassignments generally incur a small but systematic penalty in error robustness relative to the SGC.
\par
Within each type of genetic code, $D$ varies substantially across species, forming vertically dispersed clusters that reflect lineage-specific amino acid demands $f_\alpha$. Under the SGC (type 1), a clear phylogenetic signal is observed. For example, vertebrates such as \textit{Homo sapiens} (9606), \textit{Gallus gallus} (9031), and \textit{Danio rerio} (7955) cluster tightly around $D$ z-scores of approximately -3.36 to -3.19. The values in parentheses indicate the Taxonomy IDs listed in Table \ref{tab:ncbi_data}. Land plants, including \textit{Arabidopsis thaliana} (3702) and \textit{Zea mays} (4577), show a similarly constrained distribution, suggesting a conserved amino acid usage within these major kingdoms. In contrast, fungi are moderately higher and more dispersed (e.g., \textit{Saccharomyces cerevisiae} (43959), -2.84; \textit{Neurospora crassa} (367110), -3.28), indicating a different spectrum of protein demand. A notable outlier is the protozoan parasite \textit{Plasmodium falciparum} (36329, -0.09), consistent with its extreme compositional bias causing a poor match between its $f_\alpha$ and the SGC-induced $p_\alpha$. For non-SGC codes, similar phylogenetic patterns emerge. The two type 10 ciliates, \textit{Moneuplotes crassus} (5936) and \textit{Stentor coeruleus} (5963), show nearly identical $D$ values, strongly supporting a phylogenetic signal in amino acid demand. Similarly, the Mycoplasmatales/Spiroplasma group (type 4) occupies a distinct range of $D$ from -2.31 to -1.36. The ciliate codes (type 6) display a wider spread (-2.92 to -1.16), and the alternative yeast code (type 12) lies mostly between -3.14 and -1.93, with the single representative of type 26, \textit{Pachysolen tannophilus} (669874), near -2.45. This consistent clustering of closely related taxa within a given code table reinforces the idea that amino acid demand, and thus the $D$ value, is shaped by shared evolutionary history.
\par
Taken together, Figure~\ref{fig:FIG7} indicates a practical decoupling between code identity and lineage-specific proteome demand: code type discretely fixes $E$, whereas $D$ varies within each code because it is driven primarily by differences in $f_{\alpha}$ rather than by the (nearly constant) $p_{\alpha}$. Species with low $D$ are those whose amino acid usage happens to be closer to the code-imposed baseline, while higher $D$ reflects lineage-specific biases (e.g., nucleotide composition, metabolic or ecological constraints). Because our $f_{\alpha}$ estimates are derived from CDS, cross-taxon comparability is imperfect; thus, inferences about selection on $D$ should be made cautiously, with the understanding that $D$ registers both biological demand and measurement choices.
\par
In this light, we interpret $D$ as capturing compatibility between codon-imposed amino acid supply and organismal demand, complementing error robustness $E$ in a multi-objective view of code evolution. The variant codes examined here do not display a consistent shift toward lower $D$ relative to SGC-bearing lineages, providing no positive evidence that recent reassignments were primarily selected to reduce $D$. Rather, reassignments appear to have been tolerated within constraints imposed by translational robustness and lineage-specific machinery, with clade-specific $f_{\alpha}$ potentially influencing which changes were permissible. Overall, Figure~\ref{fig:FIG7} is consistent with a scenario in which $E$ imposes a strong code topological constraint, while $D$ is shaped largely within codes as proteomes diverge across lineages.

\begin{table*}
\caption{\label{tab:ncbi_data}NCBI genome dataset including species, taxonomy ID, assembly level, and coding statistics.}
\begin{ruledtabular}
\begin{tabular}{c c l c r r}
Code type & Taxonomy ID & Species & Assembly level & $N_{\text{CDS}}$ & $N_{\text{codon}}$\\
\hline
\multirow{20}{*}{1} & 400682 & \textit{Amphimedon queenslandica} & Scaffold & 23,543 & 11,541,819\\
 & 28377 & \textit{Anolis carolinensis} & Chromosome & 47,477 & 34,995,297\\
 & 3702 & \textit{Arabidopsis thaliana} & Chromosome & 48,265 & 20,890,152\\
 & 6239 & \textit{Caenorhabditis elegans} & Complete & 30,547 & 14,291,119\\
 & 7719 & \textit{Ciona intestinalis} & Chromosome & 21,099 & 13,525,781\\
 & 7955 & \textit{Danio rerio} & Chromosome & 58,095 & 41,125,855\\
 & 34168 & \textit{Diphasiastrum complanatum} & Chromosome & 69,042 & 37,073,159\\
 & 7227 & \textit{Drosophila melanogaster} & Chromosome & 30,802 & 20,409,720\\
 & 9031 & \textit{Gallus gallus} & Chromosome & 68,754 & 50,423,382\\
 & 9606 & \textit{Homo sapiens} & Chromosome & 145,439 & 99,055,406\\
 & 1480154 & \textit{Marchantia polymorpha subsp. ruderalis} & Chromosome & 20,354 & 9,025,883\\
 & 45351 & \textit{Nematostella vectensis} & Chromosome & 32,370 & 22,938,371\\
 & 367110 & \textit{Neurospora crassa} OR74A & Chromosome & 10,812 & 5,643,325\\
 & 436017 & \textit{Ostreococcus lucimarinus} CCE9901 & Complete & 7,619 & 3,080,628\\
 & 36651 & \textit{Penicillium digitatum} & Complete & 9,002 & 4,396,289\\
 & 3218 & \textit{Physcomitrium patens} & Chromosome & 48,022 & 26,956,615\\
 & 36329 & \textit{Plasmodium falciparum} 3D7 & Complete & 5,354 & 4,100,300\\
 & 559292 & \textit{Saccharomyces cerevisiae} S288C & Complete & 6,027 & 2,941,688\\
 & 8355 & \textit{Xenopus laevis} & Chromosome & 73,131 & 53,046,750\\
 & 4577 & \textit{Zea mays} & Chromosome & 57,578 & 26,343,765\\
\hline
\multirow{10}{*}{4} & 1427984 & \textit{Candidatus Hepatoplasma crinochetorum} Av & Complete & 584 & 202,071\\
 & 1609546 & \textit{endosymbiont DhMRE of Dentiscutata heterogama} & Chromosome & 766 & 207,612\\
 & 74700 & \textit{Entomoplasma freundtii} & Complete & 698 & 251,896\\
 & 2149 & \textit{Mesoplasma entomophilum} & Complete & 726 & 261,010\\
 & 265311 & \textit{Mesoplasma florum} L1 & Complete & 688 & 245,005\\
 & 941640 & \textit{Mycoplasma haemofelis} str. Langford 1 & Complete & 1,524 & 364,172\\
 & 40477 & \textit{Mycoplasma mycoides subsp. capri} & Complete & 829 & 315,558\\
 & 2133 & \textit{Spiroplasma citri} & Complete & 2,100 & 454,802\\
 & 2138 & \textit{Spiroplasma poulsonii} & Complete & 2,272 & 551,001\\
 & 214888 & \textit{Williamsoniiplasma luminosum} & Complete & 862 & 308,809\\
\hline
\multirow{10}{*}{6} & 5974 & \textit{Halteria grandinella} & Scaffold & 17,815 & 6,418,227\\
 & 28002 & \textit{Hexamita inflata} & Scaffold & 79,341 & 27,939,086\\
 & 5932 & \textit{Ichthyophthirius multifiliis} & Scaffold & 8,056 & 3,344,435\\
 & 1172189 & \textit{Oxytricha trifallax} & Contig & 810 & 193,011\\
 & 43138 & \textit{Paramecium pentaurelia} & Scaffold & 41,578 & 19,641,457\\
 & 5888 & \textit{Paramecium tetraurelia} & Scaffold & 39,642 & 18,014,858\\
 & 266149 & \textit{Pseudocohnilembus persalinus} & Scaffold & 13,179 & 7,268,791\\
 & 348837 & \textit{Spironucleus salmonicida} & Chromosome & 8,667 & 3,333,878\\
 & 5949 & \textit{Stylonychia lemnae} & Contig & 20,740 & 11,936,721\\
 & 312017 & \textit{Tetrahymena thermophila} SB210 & Scaffold & 26,996 & 16,948,603\\
\hline
\multirow{2}{*}{10} & 5936 & \textit{Moneuplotes crassus} & Contig & 29,110 & 12,561,140\\
 & 5963 & \textit{Stentor coeruleus} & Contig & 31,426 & 13,300,696\\
\hline
\multirow{10}{*}{12} & 1344418 & \textit{Ascoidea rubescens} DSM 1968 & Scaffold & 6,787 & 3,163,949\\
 & 291208 & \textit{Australozyma saopauloensis} & Complete & 4,994 & 2,551,068\\
 & 984486 & \textit{Babjeviella inositovora} NRRL Y-12698 & Scaffold & 6,398 & 2,772,846\\
 & 237561 & \textit{Candida albicans} SC5314 & Chromosome & 6,030 & 2,985,172\\
 & 498019 & \textit{Candidozyma auris} & Complete & 5,327 & 2,649,735\\
 & 36914 & \textit{Lodderomyces elongisporus} & Complete & 5,630 & 3,012,950\\
 & 2163413 & \textit{Metschnikowia aff. pulcherrima} & Complete & 5,800 & 2,910,865\\
 & 43959 & \textit{Saccharomycopsis crataegensis} & Scaffold & 6,424 & 3,297,466\\
 & 322104 & \textit{Scheffersomyces stipitis} CBS 6054 & Chromosome & 5,818 & 2,868,600\\
 & 2315449 & \textit{Yamadazyma tenuis} & Complete & 5,226 & 2,717,165\\
\hline
\multirow{5}{*}{25} & 1643353 & \textit{candidate division SR1 bacterium} Aalborg\_AAW-1 & Complete & 993 & 315,538\\
 & 1394709 & \textit{candidate division SR1 bacterium} RAAC1\_SR1\_1 & Complete & 1,059 & 357,013\\
 & 2735562 & \textit{Candidatus Absconditicoccus praedator} & Complete & 1,034 & 335,058\\
 & 1862133 & \textit{Candidatus Gracilibacteria bacterium} 28\_42\_T64 & Complete & 1,196 & 395,638\\
 & 1540872 & \textit{Candidatus Gracilibacteria bacterium} HOT-871 & Complete & 1,112 & 348,909\\
\hline
26 & 669874 & \textit{Pachysolen tannophilus} NRRL Y-2460 & Scaffold & 5,666 & 2,717,082\\
\end{tabular}
\end{ruledtabular}
\end{table*}

\section{Discussion}
\label{sec:discussion}

This study examined the origin and organizing principles of the standard genetic code (SGC) in terms of its optimization, balancing translational fidelity and amino acid diversity. The present analysis incorporates realistic mutation rates varying by codon position and type, and the amino acid composition observed in living organisms. We quantified translational error load ($E$) and compositional discrepancy ($D$) between observed amino acid usage and that implied by codon assignments--- with the explicit treatment of $D$ being a key to the analysis.
\par
Using a simulated annealing algorithm, we generated optimized genetic codes and compared them with the SGC. The results indicate that the SGC lies near a local optimum balancing error minimization and compositional alignment. Previous studies that challenged the neutral “frozen accident” view \cite{Crick68} mostly focused on minimizing translational error load \cite{Woese66, Koonin09, Massey08}. Our findings reveal an additional selective pressure: compositional matching between codon-imposed amino acid proportions and proteome usage. Overall, these result suggest that the genetic code evolved under selection balancing error load and compositional discrepancy.
\par
The SGC is exceptionally rare with respect to both $E$ and $D$, with the improbability of its error load particularly striking. Local variation analyses show that the SGC reaches a local optimum for small values of the balancing parameter $\eta$ in the loss function $L = E + \eta D$. From $5.8\times10^7$ random codes, $E$ and $D$ are found to be uncorrelated; although $D$ is less statistically extreme than $E$, its deviation remains significant. These results support the view that the genetic code evolved under selection minimizing both error load and compositional discrepancy, while the precise evolutionary weighting $\eta$ remains uncertain.

\par
This work provides a simplified model of the evolutionary trajectory of the genetic code. The loss function defined here serves dual purposes: quantifying the optimization of the present genetic code and establishing a framework for simulating the evolution from ancestral to contemporary forms. Analysis of local variations around the SGC indicates that preservation of synonymous codon blocks is critical for maintaining stability under specific environmental conditions.
\par
Several limitations constrain this study. First, although simulated annealing provides an effective optimization algorithm for genetic codes, its trajectory does not directly correspond to natural evolutionary processes. The exponential temperature decrease in simulated annealing---transitioning from exploration to exploitation---may not accurately represent complex biological evolution. Similarly, the balancing parameter $\eta$ likely varied throughout evolutionary history rather than remaining constant as assumed here. Most fundamentally, the billion-year timescale of genetic code evolution precludes accurate reconstruction of the environmental factors that shaped its development.
\par
Second, the amino acid Kullback-Leibler divergence term quantifies the discrepancy between the natural frequency $f_{\alpha}$ and the code frequency $p_{\alpha}$. Our study assumes that the amino acid requirements of an organism are represented by $f_{\alpha}$, and that the genetic code evolved under selective pressures to bring $p_{\alpha}$ closer to $f_{\alpha}$. However, we computed $f_{\alpha}$ based on coding DNA sequence (CDS) data from genomic datasets, which introduces a fundamental interpretive challenge. If nucleotide sequences were entirely random with equal base frequencies, the empirically calculated $f_{\alpha}$ would exactly match $p_{\alpha}$ by mathematical construction. This creates a circular dependency where the observed similarity between $f_{\alpha}$ and $p_{\alpha}$ could arise purely from the mathematical relationship between codon usage and amino acid frequencies, rather than from evolutionary optimization of the genetic code itself. Consequently, it becomes difficult to distinguish whether apparent code optimization reflects genuine selective pressures or simply emerges as an inevitable consequence of the statistical relationship between nucleotide composition and amino acid frequencies.
\par
Third, while we addressed variant genetic codes to the extent possible, our study was constrained by data availability and focused primarily on the standard genetic code and its close variants. Notably, mitochondrial and chloroplast codes were excluded from our analysis due to their distinct evolutionary histories \cite{Osawa92} and specialized functional contexts, which may involve different selective pressures compared to nuclear translation systems. Additionally, several recently discovered variant codes in specific bacterial and archaeal lineages could not be included due to limited genomic data availability. Although our focus on nuclear codes was motivated by their central role in cellular protein synthesis and the abundance of comparative genomic data, incorporating a broader range of variant codes could provide additional insights into the generalizability of our optimization framework.
\par
Future studies should aim to address these limitations through both experimental and computational approaches. While direct manipulation of genetic codes remains technically challenging, targeted experiments using engineered tRNA variants or codon reassignment systems \cite{Lajoie13, Chin17} could provide empirical validation of our error minimization predictions by measuring translational fidelity and loss consequences. From a computational perspective, evolutionary simulations should expand beyond simple error minimization models to incorporate environmental factors such as temperature, chemical composition, and metabolic constraints that likely influenced early genetic code evolution. Such models could help resolve the circular dependency issue by explicitly modeling the co-evolution of nucleotide composition and amino acid usage under realistic selective pressures. Additionally, systematic comparative analysis of natural variant codes---particularly those with well-documented codon reassignments in mitochondria and specific microbial lineages---could provide independent validation of our optimization framework while revealing how different evolutionary contexts shape code structure. These complementary approaches would strengthen the empirical foundation for understanding genetic code evolution beyond the theoretical predictions presented here.

\section*{Acknowledgment}
This work was supported by the Creative-Pioneering Researchers Program through Seoul National University, and the National Research Foundation of Korea (NRF) grant (Grant No. 2022R1A2C1006871), and the ICTP through the Associates Programme (2020-2025) (J.J.).

\appendix*

\section{Threshold value of the third transition weight}
\label{sec:appendix}

Since all codons are robust to third position transition (third transition) mutations, any local variation that alters the encoded amino acid necessarily disrupts this robustness. Assuming that double or higher-order mutations are negligible ($c^2 \ll c$), the error load is affected only by the amino acid distance between the varied codon and its neighboring codons that differ by a point mutation. Let $\alpha$ represent the codon altered by the local variation. Since each of the three codon positions allows one transition and two transversion mutations, $\alpha$ has up to nine neighboring codons. This number may decrease to eight or seven if any of the neighboring codons correspond to stop codons. Let $\aleph$ denote the set of neighboring codons, excluding those that arise from third-position transition mutations or correspond to stop codons. Let $w_i$ represent the mutation weight associated with each codon $i$ in this set.
\par
In the case of absolute error, the amino acid distance is determined by the difference in polar requirement values $\phi$ between the amino acids specified by two codons. Let the polar requirement of the amino acid specified by codon $\alpha$ change from $\phi_\alpha$ to $\phi_\alpha^\ast$ due to local variation. Then, the corresponding change in error load, $\Delta E$, can be approximated as follows:

\begin{eqnarray} \label{eq:A.1}
\Delta E\left(\phi_\alpha\rightarrow\phi_{\alpha}^{\ast}\right)&\approx&\sum_{i}w_{i}\big(\vert\phi_i-\phi_{\alpha}^{\ast}\vert-\vert\phi_i-\phi_\alpha\vert\big)\nonumber \\ &=&\sum_{i\in\aleph}w_{i}\big(\vert\phi_i-\phi_{\alpha}^{\ast}\vert-\vert\phi_i-\phi_\alpha\vert\big)\nonumber \\
 &&+w_{\mathrm{tt}}\big(\vert\phi_{\mathrm{tt}}-\phi_{\alpha}^{\ast}\vert-\vert\phi_{\mathrm{tt}}-\phi_\alpha\vert\big).\nonumber \\
\end{eqnarray}

Let $\delta \equiv \phi_{\alpha}^\ast - \phi_\alpha$ and $\tilde{\phi}_i \equiv \phi_i - \phi_\alpha$. Then, $\Delta E$ becomes:

\begin{eqnarray} \label{eq:A.2}
\Delta E\left(\phi_\alpha\rightarrow\phi_{\alpha}^{\ast}\right)&\approx&\sum_{i\in\aleph}w_{i}\left(\big\vert\tilde{\phi}_i-\delta\big\vert-\big\vert\tilde{\phi}_i\big\vert\right)\nonumber \\
 &&+ w_{\mathrm{tt}}\left(\big\vert\tilde{\phi}_{\mathrm{tt}}-\delta\big\vert-\big\vert\tilde{\phi}_{\mathrm{tt}}\big\vert\right).
\end{eqnarray}

For the SGC to qualify as a local optimum, it must satisfy $\Delta E \leq 0$ for all $\alpha$. Since robustness to third transition mutations is established across all codons, we have $\tilde{\phi}_{\mathrm{tt}} = 0$. Given that $w_{\mathrm{tt}}$ is the largest among the six weights, the threshold value $w_{\mathrm{tt}}^\ast$ at which the SGC becomes a local optimum is determined as follows:

\begin{equation} \label{eq:A.3}
w_{\mathrm{tt}}^\ast\approx\max\left\{\frac{1}{\vert\delta\vert}\sum_{i\in\aleph}w_{i}\left(\big\vert\tilde{\phi}_i\big\vert-\big\vert\tilde{\phi}_i-\delta\big\vert\right)\right\}.
\end{equation}

From Eq.~\eqref{eq:A.3}, a necessary condition for the expression inside the braces on the right-hand side to be maximal is that, for all $i$, either $\phi_i \leq \phi_{\alpha}^{\ast} < \phi_\alpha$ or $\phi_i \geq \phi_{\alpha}^{\ast} > \phi_\alpha$. This implies that $\phi_\alpha$ must be either the maximum or minimum among all $\phi_i$, and that $\phi_{\alpha}^{\ast}$ is the second-largest or second-smallest, respectively. Hence, under this condition, we have $\big\vert\tilde{\phi}_i\big\vert-\big\vert\tilde{\phi}_i-\delta\big\vert=\vert\delta\vert$, and thus

\begin{equation} \label{eq:A.4}
w_{\mathrm{tt}}^\ast \approx \max\left\{\sum_{i\in\aleph} w_i \right\}.
\end{equation}

Analyzing the structure of the SGC reveals that the codons UGU and UGC, which encode cysteine (Cys, $\phi = 4.3$) and exhibit the lowest polar requirement, are connected via a third position transversion mutation ($w_3^{\mathrm{tv}} = 1$) to the neighboring stop codon UGA. In contrast, the codons GAA and GAG, which encode glutamic acid (Glu, $\phi = 13.6$) and exhibit the highest polar requirement, are connected via a first position transversion mutation ($w_1^{\mathrm{tv}} = 0.5$) to the neighboring stop codons UAA and UAG. Accordingly, the threshold value is estimated as $w_{\mathrm{tt}}^\ast \approx 4.2$ under the assumption of local variation in GAA or GAG.
\par
The value of $w_{\mathrm{tt}}^\ast$ can also be derived analogously in the case of squared error. The change in error load, $\Delta E$, takes the same form as that in Eq.~\eqref{eq:A.1} for absolute error, except that the absolute difference in polar requirement is replaced by the square of the difference:

\begin{eqnarray} \label{eq:A.5}
\Delta E\left(\phi_\alpha\rightarrow\phi_{\alpha}^{\ast}\right)&\approx&\sum_{i}w_{i}\left[\left(\phi_i-\phi_{\alpha}^{\ast}\right)^2-\left(\phi_i-\phi_\alpha\right)^2\right]\nonumber \\ &=&\sum_{i\in\aleph}w_{i}\left[\big(\tilde{\phi}_i-\delta\big)^2\right]\nonumber \\&&+w_{\mathrm{tt}}\left[\big(\tilde{\phi}_{\mathrm{tt}}-\delta\big)^2-\big(\tilde{\phi}_{\mathrm{tt}}\big)^2\right]\nonumber \\ &=&\sum_{i\in\aleph}w_{i}\delta\big(\delta-2\tilde{\phi}_i\big)+w_{\mathrm{tt}}\delta\big(\delta-2\tilde{\phi}_{\mathrm{tt}}\big).\nonumber \\
\end{eqnarray}

Similarly, since $\tilde{\phi}_{\mathrm{tt}} = 0$, the threshold value $w_{\mathrm{tt}}^\ast$ is given by:

\begin{eqnarray} \label{eq:A.6}
w_{\mathrm{tt}}^{\ast}&\approx&\max\left\{\frac{1}{\delta^2}\sum_{i\in\aleph}w_{i}\delta\big(2\tilde{\phi}_i-\delta\big)\right\}\nonumber \\
 &=&\max\left\{\frac{2}{\delta}\sum_{i\in\aleph}w_{i}\tilde{\phi}_i-\sum_{i\in\aleph}w_i\right\}.\nonumber \\
\end{eqnarray}

The expression inside the brackets is maximized when the signs of $\delta$ and $\tilde{\phi}_i$ are identical for all $i$, $\vert\delta\vert$ is minimized, and $\big\vert\tilde{\phi}_i\big\vert$ is maximized. Given the values of polar requirement, the minimum possible value of $\vert\delta\vert$ is 0.1. To maximize $\big\vert\tilde{\phi}_i\big\vert$, $\phi_\alpha$ must be either the maximum or minimum among all polar requirement values. In the SGC, this condition is satisfied by the codons UGU and UGC, which specify the amino acid with the lowest polar requirement ($\phi=4.3$) and yield $\vert\delta\vert = 0.1$. Accordingly, under the assumption of local variation in these codons, the resulting threshold value is calculated as $w_{\mathrm{tt}}^\ast \approx 214.1$.
\par
Having analyzed $w_{\mathrm{tt}}^\ast$ under absolute and squared error, we now generalize the formulation to the case where the amino acid distance is defined as $\vert\Delta \phi\vert^n$. As in the previous two cases, we assume that the codons specifying the amino acid with the maximal or minimal $\phi_\alpha$ undergo local variation such that $\phi_\alpha^\ast$ becomes the second-largest or second-smallest value. Under this assumption, and applying the binomial theorem, the change in error load $\Delta E$ can be expressed as:

\begin{eqnarray} \label{eq:A.7}
\Delta E\left(\phi_\alpha\rightarrow\phi_{\alpha}^{\ast}\right)&\approx&\sum_{i}w_{i}\sum_{k=1}^{n}(-1)^{k}\binom{n}{k}\big\vert\tilde{\phi}_i\big\vert^{n-k}\vert\delta\vert^k \nonumber \\
 &=&\sum_{i\in\aleph}w_i\sum_{k=1}^{n}(-1)^{k}\binom{n}{k}\big\vert\tilde{\phi}_i\big\vert^{n-k}\vert\delta\vert^k \nonumber \\[5pt] &&+w_{\mathrm{tt}}\vert\delta\vert^n.
\end{eqnarray}

Setting $\Delta E = 0$, the threshold value $w_{\mathrm{tt}}^\ast$ is given by:

\begin{equation} \label{eq:A.8}
w_{\mathrm{tt}}^\ast\approx\max\left\{\sum_{i\in\aleph}w_i\sum_{k=1}^{n}(-1)^{k-1}\binom{n}{k}\bigg\vert\frac{\tilde{\phi}_i}{\delta}\bigg\vert^{n-k}\right\}.
\end{equation}

Assuming that $\vert\delta\vert \ll \big\vert\tilde{\phi}_i\big\vert$, the threshold value $w{\mathrm{tt}}^\ast$ can be approximated by retaining only the leading-order term of the binomial expansion:

\begin{equation} \label{eq:A.9}
w_{\mathrm{tt}}^\ast\sim\max\left\{n\sum_{i\in\aleph}w_i\bigg\vert\frac{\tilde{\phi}_i}{\delta}\bigg\vert^{n-1}\right\}.
\end{equation}

Since the minimum possible value of $\vert\delta\vert$ is 0.1, the threshold value $w_{\mathrm{tt}}^\ast$ increases rapidly as the exponent $n$ increases. Moreover, UGU and UGC are the only codons that satisfy $\vert\delta\vert = 0.1$ and either maximize or minimize $\big\vert\tilde{\phi}_i\big\vert$. Therefore, for $n \geq 2$, the codons that determine the threshold value for the third-position transition weight are uniquely fixed as UGU and UGC.

\nocite{*}

\bibliography{apssamp}

\providecommand{\noopsort}[1]{}\providecommand{\singleletter}[1]{#1}%
\begin{thebibliography}{60}%
\makeatletter
\providecommand \@ifxundefined [1]{%
 \@ifx{#1\undefined}
}%
\providecommand \@ifnum [1]{%
 \ifnum #1\expandafter \@firstoftwo
 \else \expandafter \@secondoftwo
 \fi
}%
\providecommand \@ifx [1]{%
 \ifx #1\expandafter \@firstoftwo
 \else \expandafter \@secondoftwo
 \fi
}%
\providecommand \natexlab [1]{#1}%
\providecommand \enquote  [1]{``#1''}%
\providecommand \bibnamefont  [1]{#1}%
\providecommand \bibfnamefont [1]{#1}%
\providecommand \citenamefont [1]{#1}%
\providecommand \href@noop [0]{\@secondoftwo}%
\providecommand \href [0]{\begingroup \@sanitize@url \@href}%
\providecommand \@href[1]{\@@startlink{#1}\@@href}%
\providecommand \@@href[1]{\endgroup#1\@@endlink}%
\providecommand \@sanitize@url [0]{\catcode `\\12\catcode `\$12\catcode `\&12\catcode `\#12\catcode `\^12\catcode `\_12\catcode `\%12\relax}%
\providecommand \@@startlink[1]{}%
\providecommand \@@endlink[0]{}%
\providecommand \url  [0]{\begingroup\@sanitize@url \@url }%
\providecommand \@url [1]{\endgroup\@href {#1}{\urlprefix }}%
\providecommand \urlprefix  [0]{URL }%
\providecommand \Eprint [0]{\href }%
\providecommand \doibase [0]{https://doi.org/}%
\providecommand \selectlanguage [0]{\@gobble}%
\providecommand \bibinfo  [0]{\@secondoftwo}%
\providecommand \bibfield  [0]{\@secondoftwo}%
\providecommand \translation [1]{[#1]}%
\providecommand \BibitemOpen [0]{}%
\providecommand \bibitemStop [0]{}%
\providecommand \bibitemNoStop [0]{.\EOS\space}%
\providecommand \EOS [0]{\spacefactor3000\relax}%
\providecommand \BibitemShut  [1]{\csname bibitem#1\endcsname}%
\let\auto@bib@innerbib\@empty
\bibitem [{\citenamefont {Knight}\ \emph {et~al.}(2001)\citenamefont {Knight}, \citenamefont {Freeland},\ and\ \citenamefont {Landweber}}]{Knight01}%
  \BibitemOpen
  \bibfield  {author} {\bibinfo {author} {\bibfnamefont {R.~D.}\ \bibnamefont {Knight}}, \bibinfo {author} {\bibfnamefont {S.~J.}\ \bibnamefont {Freeland}},\ and\ \bibinfo {author} {\bibfnamefont {L.~F.}\ \bibnamefont {Landweber}},\ }\bibfield  {title} {\bibinfo {title} {Rewiring the keyboard: evolvability of the genetic code},\ }\href {https://doi.org/10.1038/35047500} {\bibfield  {journal} {\bibinfo  {journal} {Nat. Rev. Genet.}\ }\textbf {\bibinfo {volume} {2}},\ \bibinfo {pages} {49} (\bibinfo {year} {2001})}\BibitemShut {NoStop}%
\bibitem [{\citenamefont {Osawa}(1992)}]{Osawa92}%
  \BibitemOpen
  \bibfield  {author} {\bibinfo {author} {\bibfnamefont {S.}~\bibnamefont {Osawa}},\ }\bibfield  {title} {\bibinfo {title} {Recent evidence for evolution of the genetic code},\ }\href {https://doi.org/10.1128/mr.56.1.229-264.1992} {\bibfield  {journal} {\bibinfo  {journal} {Microbiological reviews}\ }\textbf {\bibinfo {volume} {56}},\ \bibinfo {pages} {229} (\bibinfo {year} {1992})}\BibitemShut {NoStop}%
\bibitem [{\citenamefont {Błażej}\ \emph {et~al.}(2018)\citenamefont {Błażej}, \citenamefont {Wnętrzak}, \citenamefont {Mackiewicz},\ and\ \citenamefont {Mackiewicz}}]{Blazej18}%
  \BibitemOpen
  \bibfield  {author} {\bibinfo {author} {\bibfnamefont {P.}~\bibnamefont {Błażej}}, \bibinfo {author} {\bibfnamefont {M.}~\bibnamefont {Wnętrzak}}, \bibinfo {author} {\bibfnamefont {D.}~\bibnamefont {Mackiewicz}},\ and\ \bibinfo {author} {\bibfnamefont {P.}~\bibnamefont {Mackiewicz}},\ }\bibfield  {title} {\bibinfo {title} {Optimization of the standard genetic code according to three codon positions using an evolutionary algorithm},\ }\href {https://doi.org/10.1371/journal.pone.0205450} {\bibfield  {journal} {\bibinfo  {journal} {Plos one}\ }\textbf {\bibinfo {volume} {13}},\ \bibinfo {pages} {e0201715} (\bibinfo {year} {2018})}\BibitemShut {NoStop}%
\bibitem [{\citenamefont {Sch{\"o}nauer}\ and\ \citenamefont {Clote}(1997)}]{Schnauer97}%
  \BibitemOpen
  \bibfield  {author} {\bibinfo {author} {\bibfnamefont {S.}~\bibnamefont {Sch{\"o}nauer}}\ and\ \bibinfo {author} {\bibfnamefont {P.}~\bibnamefont {Clote}},\ }\bibfield  {title} {\bibinfo {title} {How optimal is the genetic code?},\ }in\ \href@noop {} {\emph {\bibinfo {booktitle} {German Conference on Bioinformatics}}}\ (\bibinfo {year} {1997})\BibitemShut {NoStop}%
\bibitem [{\citenamefont {Crick}(1968)}]{Crick68}%
  \BibitemOpen
  \bibfield  {author} {\bibinfo {author} {\bibfnamefont {F.~H.}\ \bibnamefont {Crick}},\ }\bibfield  {title} {\bibinfo {title} {The origin of the genetic code},\ }\href {https://doi.org/10.1016/0022-2836(68)90392-6} {\bibfield  {journal} {\bibinfo  {journal} {Journal of molecular biology}\ }\textbf {\bibinfo {volume} {120}},\ \bibinfo {pages} {367} (\bibinfo {year} {1968})}\BibitemShut {NoStop}%
\bibitem [{\citenamefont {Di~Giulio}(2005)}]{DiGiulio2005}%
  \BibitemOpen
  \bibfield  {author} {\bibinfo {author} {\bibfnamefont {M.}~\bibnamefont {Di~Giulio}},\ }\bibfield  {title} {\bibinfo {title} {The origin of the genetic code: theories and their relationships, a review},\ }\href {https://doi.org/10.1016/j.biosystems.2004.11.005} {\bibfield  {journal} {\bibinfo  {journal} {BioSystems}\ }\textbf {\bibinfo {volume} {80}},\ \bibinfo {pages} {175} (\bibinfo {year} {2005})}\BibitemShut {NoStop}%
\bibitem [{\citenamefont {Tlusty}(2010)}]{Tlusty10}%
  \BibitemOpen
  \bibfield  {author} {\bibinfo {author} {\bibfnamefont {T.}~\bibnamefont {Tlusty}},\ }\bibfield  {title} {\bibinfo {title} {A colorful origin for the genetic code: Information theory, statistical mechanics and the emergence of molecular codes},\ }\href {https://doi.org/10.1016/j.plrev.2010.06.002} {\bibfield  {journal} {\bibinfo  {journal} {Physics of life reviews}\ }\textbf {\bibinfo {volume} {7}},\ \bibinfo {pages} {362} (\bibinfo {year} {2010})}\BibitemShut {NoStop}%
\bibitem [{\citenamefont {Sonneborn}(1965)}]{Sonneborn1965}%
  \BibitemOpen
  \bibfield  {author} {\bibinfo {author} {\bibfnamefont {T.}~\bibnamefont {Sonneborn}},\ }\bibfield  {title} {\bibinfo {title} {Degeneracy of the genetic code: extent, nature, and genetic implications},\ }in\ \href@noop {} {\emph {\bibinfo {booktitle} {Evolving genes and proteins}}}\ (\bibinfo  {publisher} {Elsevier},\ \bibinfo {year} {1965})\ pp.\ \bibinfo {pages} {377--397}\BibitemShut {NoStop}%
\bibitem [{\citenamefont {Sella}\ and\ \citenamefont {Ardell}(2006)}]{Sella2006}%
  \BibitemOpen
  \bibfield  {author} {\bibinfo {author} {\bibfnamefont {G.}~\bibnamefont {Sella}}\ and\ \bibinfo {author} {\bibfnamefont {D.~H.}\ \bibnamefont {Ardell}},\ }\bibfield  {title} {\bibinfo {title} {The coevolution of genes and genetic codes: Crick’s frozen accident revisited},\ }\href@noop {} {\bibfield  {journal} {\bibinfo  {journal} {Journal of molecular evolution}\ }\textbf {\bibinfo {volume} {63}},\ \bibinfo {pages} {297} (\bibinfo {year} {2006})}\BibitemShut {NoStop}%
\bibitem [{\citenamefont {Woese}\ \emph {et~al.}(1966)\citenamefont {Woese}, \citenamefont {Dugre}, \citenamefont {Saxinger},\ and\ \citenamefont {Dugre}}]{Woese66}%
  \BibitemOpen
  \bibfield  {author} {\bibinfo {author} {\bibfnamefont {C.~R.}\ \bibnamefont {Woese}}, \bibinfo {author} {\bibfnamefont {D.~H.}\ \bibnamefont {Dugre}}, \bibinfo {author} {\bibfnamefont {W.~C.}\ \bibnamefont {Saxinger}},\ and\ \bibinfo {author} {\bibfnamefont {S.~A.}\ \bibnamefont {Dugre}},\ }\bibfield  {title} {\bibinfo {title} {The molecular basis for the genetic code},\ }\href {https://doi.org/10.1073/pnas.55.4.966} {\bibfield  {journal} {\bibinfo  {journal} {Proceedings of the National Academy of Sciences}\ }\textbf {\bibinfo {volume} {55}},\ \bibinfo {pages} {966} (\bibinfo {year} {1966})}\BibitemShut {NoStop}%
\bibitem [{\citenamefont {Vetsigian}\ \emph {et~al.}(2006)\citenamefont {Vetsigian}, \citenamefont {Woese},\ and\ \citenamefont {Goldenfeld}}]{Vestigian2006}%
  \BibitemOpen
  \bibfield  {author} {\bibinfo {author} {\bibfnamefont {K.}~\bibnamefont {Vetsigian}}, \bibinfo {author} {\bibfnamefont {C.}~\bibnamefont {Woese}},\ and\ \bibinfo {author} {\bibfnamefont {N.}~\bibnamefont {Goldenfeld}},\ }\bibfield  {title} {\bibinfo {title} {Collective evolution and the genetic code},\ }\href {https://doi.org/10.1073/pnas.0603780103} {\bibfield  {journal} {\bibinfo  {journal} {Proceedings of the National Academy of Sciences}\ }\textbf {\bibinfo {volume} {103}},\ \bibinfo {pages} {10696} (\bibinfo {year} {2006})}\BibitemShut {NoStop}%
\bibitem [{\citenamefont {Gamow}(1954)}]{Gamow54}%
  \BibitemOpen
  \bibfield  {author} {\bibinfo {author} {\bibfnamefont {G.}~\bibnamefont {Gamow}},\ }\bibfield  {title} {\bibinfo {title} {Possible relation between deoxyribonucleic acid and protein structures},\ }\href {https://doi.org/10.1038/173318a0} {\bibfield  {journal} {\bibinfo  {journal} {Nature}\ }\textbf {\bibinfo {volume} {173}},\ \bibinfo {pages} {318} (\bibinfo {year} {1954})}\BibitemShut {NoStop}%
\bibitem [{\citenamefont {Dunnill}(1966)}]{Dunnill1966}%
  \BibitemOpen
  \bibfield  {author} {\bibinfo {author} {\bibfnamefont {P.}~\bibnamefont {Dunnill}},\ }\bibfield  {title} {\bibinfo {title} {Triplet nucleotide-amino-acid pairing; a stereochemical basis for the division between protein and non-protein amino-acids},\ }\href {https://doi.org/10.1038/2101267a0} {\bibfield  {journal} {\bibinfo  {journal} {Nature}\ }\textbf {\bibinfo {volume} {210}},\ \bibinfo {pages} {1267} (\bibinfo {year} {1966})}\BibitemShut {NoStop}%
\bibitem [{\citenamefont {Yarus}\ and\ \citenamefont {Christian}(1989)}]{Yarus1989}%
  \BibitemOpen
  \bibfield  {author} {\bibinfo {author} {\bibfnamefont {M.}~\bibnamefont {Yarus}}\ and\ \bibinfo {author} {\bibfnamefont {E.~L.}\ \bibnamefont {Christian}},\ }\bibfield  {title} {\bibinfo {title} {Genetic code origins},\ }\href {https://doi.org/10.1038/342349b0} {\bibfield  {journal} {\bibinfo  {journal} {Nature}\ }\textbf {\bibinfo {volume} {342}},\ \bibinfo {pages} {349} (\bibinfo {year} {1989})}\BibitemShut {NoStop}%
\bibitem [{\citenamefont {Rowe}(1994)}]{Rowe94}%
  \BibitemOpen
  \bibfield  {author} {\bibinfo {author} {\bibfnamefont {G.}~\bibnamefont {Rowe}},\ }\href@noop {} {\emph {\bibinfo {title} {Theoretical Models in Biology: The Origin of Life, the Immune System, and the Brain}}}\ (\bibinfo  {publisher} {Oxford University Press},\ \bibinfo {year} {1994})\BibitemShut {NoStop}%
\bibitem [{\citenamefont {Knight}\ \emph {et~al.}(1999)\citenamefont {Knight}, \citenamefont {Freeland},\ and\ \citenamefont {Landweber}}]{Knight1999}%
  \BibitemOpen
  \bibfield  {author} {\bibinfo {author} {\bibfnamefont {R.~D.}\ \bibnamefont {Knight}}, \bibinfo {author} {\bibfnamefont {S.~J.}\ \bibnamefont {Freeland}},\ and\ \bibinfo {author} {\bibfnamefont {L.~F.}\ \bibnamefont {Landweber}},\ }\bibfield  {title} {\bibinfo {title} {Selection, history and chemistry: The three faces of the genetic code},\ }\href {https://doi.org/10.1016/s0968-0004(99)01412-2} {\bibfield  {journal} {\bibinfo  {journal} {Trends in Biochemical Sciences}\ }\textbf {\bibinfo {volume} {24}},\ \bibinfo {pages} {241} (\bibinfo {year} {1999})}\BibitemShut {NoStop}%
\bibitem [{\citenamefont {Yarus}\ \emph {et~al.}(2005)\citenamefont {Yarus}, \citenamefont {Caporaso},\ and\ \citenamefont {Knight}}]{Yarus2005}%
  \BibitemOpen
  \bibfield  {author} {\bibinfo {author} {\bibfnamefont {M.}~\bibnamefont {Yarus}}, \bibinfo {author} {\bibfnamefont {J.~G.}\ \bibnamefont {Caporaso}},\ and\ \bibinfo {author} {\bibfnamefont {R.}~\bibnamefont {Knight}},\ }\bibfield  {title} {\bibinfo {title} {Origins of the genetic code: The escaped triplet theory},\ }\href {https://doi.org/10.1146/annurev.biochem.74.082803.133119} {\bibfield  {journal} {\bibinfo  {journal} {Annual Review of Biochemistry}\ }\textbf {\bibinfo {volume} {74}},\ \bibinfo {pages} {179} (\bibinfo {year} {2005})}\BibitemShut {NoStop}%
\bibitem [{\citenamefont {Johnson}\ and\ \citenamefont {Wang}(2010)}]{Johnson2010}%
  \BibitemOpen
  \bibfield  {author} {\bibinfo {author} {\bibfnamefont {D.~B.~F.}\ \bibnamefont {Johnson}}\ and\ \bibinfo {author} {\bibfnamefont {L.}~\bibnamefont {Wang}},\ }\bibfield  {title} {\bibinfo {title} {Imprints of the genetic code in the ribosome},\ }\href {https://doi.org/10.1073/pnas.1000704107} {\bibfield  {journal} {\bibinfo  {journal} {Proceedings of the National Academy of Sciences of the United States of America}\ }\textbf {\bibinfo {volume} {107}},\ \bibinfo {pages} {8298} (\bibinfo {year} {2010})}\BibitemShut {NoStop}%
\bibitem [{\citenamefont {Goldberg}\ and\ \citenamefont {Wittes}(1966)}]{Goldberg1966}%
  \BibitemOpen
  \bibfield  {author} {\bibinfo {author} {\bibfnamefont {A.~L.}\ \bibnamefont {Goldberg}}\ and\ \bibinfo {author} {\bibfnamefont {R.~E.}\ \bibnamefont {Wittes}},\ }\bibfield  {title} {\bibinfo {title} {Genetic code: Aspects of organization},\ }\href {https://doi.org/10.1126/science.153.3734.420} {\bibfield  {journal} {\bibinfo  {journal} {Science}\ }\textbf {\bibinfo {volume} {153}},\ \bibinfo {pages} {420} (\bibinfo {year} {1966})}\BibitemShut {NoStop}%
\bibitem [{\citenamefont {Alff-Steinberger}(1969)}]{Alff-Steinbeger69}%
  \BibitemOpen
  \bibfield  {author} {\bibinfo {author} {\bibfnamefont {C.}~\bibnamefont {Alff-Steinberger}},\ }\bibfield  {title} {\bibinfo {title} {The genetic code and error transmission},\ }\href {https://doi.org/10.1073/pnas.64.2.584} {\bibfield  {journal} {\bibinfo  {journal} {Proceedings of the National Academy of Sciences}\ }\textbf {\bibinfo {volume} {64}},\ \bibinfo {pages} {584} (\bibinfo {year} {1969})}\BibitemShut {NoStop}%
\bibitem [{\citenamefont {Rumer}(2016)}]{Rumer2016trans}%
  \BibitemOpen
  \bibfield  {author} {\bibinfo {author} {\bibfnamefont {Y.~B.}\ \bibnamefont {Rumer}},\ }\bibfield  {title} {\bibinfo {title} {Translation of `systematization of codons in the genetic code [i]' by yu. b. rumer (1966)},\ }\href {https://doi.org/10.1098/rsta.2015.0446} {\bibfield  {journal} {\bibinfo  {journal} {Philosophical Transactions of the Royal Society A: Mathematical, Physical and Engineering Sciences}\ }\textbf {\bibinfo {volume} {374}},\ \bibinfo {pages} {20150446} (\bibinfo {year} {2016})},\ \bibinfo {note} {first published in Russian in 1966}\BibitemShut {NoStop}%
\bibitem [{\citenamefont {Tlusty}(2007)}]{Tlusty2007}%
  \BibitemOpen
  \bibfield  {author} {\bibinfo {author} {\bibfnamefont {T.}~\bibnamefont {Tlusty}},\ }\bibfield  {title} {\bibinfo {title} {A model for the emergence of the genetic code as a transition in a noisy information channel},\ }\href {https://doi.org/10.1016/j.jtbi.2007.07.029} {\bibfield  {journal} {\bibinfo  {journal} {Journal of Theoretical Biology}\ }\textbf {\bibinfo {volume} {249}},\ \bibinfo {pages} {331} (\bibinfo {year} {2007})}\BibitemShut {NoStop}%
\bibitem [{\citenamefont {Massey}(2008)}]{Massey08}%
  \BibitemOpen
  \bibfield  {author} {\bibinfo {author} {\bibfnamefont {S.~E.}\ \bibnamefont {Massey}},\ }\bibfield  {title} {\bibinfo {title} {A neutral origin for error minimization in the genetic code},\ }\href {https://doi.org/10.1007/s00239-008-9167-4} {\bibfield  {journal} {\bibinfo  {journal} {Journal of molecular evolution}\ }\textbf {\bibinfo {volume} {67}},\ \bibinfo {pages} {510} (\bibinfo {year} {2008})}\BibitemShut {NoStop}%
\bibitem [{\citenamefont {Koonin}\ and\ \citenamefont {Novozhilov}(2009)}]{Koonin09}%
  \BibitemOpen
  \bibfield  {author} {\bibinfo {author} {\bibfnamefont {E.~V.}\ \bibnamefont {Koonin}}\ and\ \bibinfo {author} {\bibfnamefont {A.~S.}\ \bibnamefont {Novozhilov}},\ }\bibfield  {title} {\bibinfo {title} {Origin and evolution of the genetic code: the universal enigma},\ }\href {https://doi.org/10.1002/iub.146} {\bibfield  {journal} {\bibinfo  {journal} {IUBMB life}\ }\textbf {\bibinfo {volume} {61}},\ \bibinfo {pages} {99} (\bibinfo {year} {2009})}\BibitemShut {NoStop}%
\bibitem [{\citenamefont {Haig}\ and\ \citenamefont {Hurst}(1991)}]{Haig91}%
  \BibitemOpen
  \bibfield  {author} {\bibinfo {author} {\bibfnamefont {D.}~\bibnamefont {Haig}}\ and\ \bibinfo {author} {\bibfnamefont {L.~D.}\ \bibnamefont {Hurst}},\ }\bibfield  {title} {\bibinfo {title} {A quantitative measure of error minimization in the genetic code},\ }\href {https://doi.org/10.1007/BF02103132} {\bibfield  {journal} {\bibinfo  {journal} {Journal of molecular evolution}\ }\textbf {\bibinfo {volume} {33}},\ \bibinfo {pages} {412} (\bibinfo {year} {1991})}\BibitemShut {NoStop}%
\bibitem [{\citenamefont {Freeland}\ and\ \citenamefont {Hurst}(1998{\natexlab{a}})}]{Freeland98}%
  \BibitemOpen
  \bibfield  {author} {\bibinfo {author} {\bibfnamefont {S.~J.}\ \bibnamefont {Freeland}}\ and\ \bibinfo {author} {\bibfnamefont {L.~D.}\ \bibnamefont {Hurst}},\ }\bibfield  {title} {\bibinfo {title} {The genetic code is one in a million},\ }\href {https://doi.org/10.1007/PL00006381} {\bibfield  {journal} {\bibinfo  {journal} {Journal of molecular evolution}\ }\textbf {\bibinfo {volume} {47}},\ \bibinfo {pages} {238} (\bibinfo {year} {1998}{\natexlab{a}})}\BibitemShut {NoStop}%
\bibitem [{\citenamefont {Freeland}\ and\ \citenamefont {Hurst}(1998{\natexlab{b}})}]{Freeland98_2}%
  \BibitemOpen
  \bibfield  {author} {\bibinfo {author} {\bibfnamefont {S.~J.}\ \bibnamefont {Freeland}}\ and\ \bibinfo {author} {\bibfnamefont {L.~D.}\ \bibnamefont {Hurst}},\ }\bibfield  {title} {\bibinfo {title} {Load minimization of the genetic code: history does not explain the pattern},\ }\href {https://doi.org/10.1098/rspb.1998.0547} {\bibfield  {journal} {\bibinfo  {journal} {Proceedings of the Royal Society B: Biological Sciences}\ }\textbf {\bibinfo {volume} {265}},\ \bibinfo {pages} {2111} (\bibinfo {year} {1998}{\natexlab{b}})}\BibitemShut {NoStop}%
\bibitem [{\citenamefont {Ardell}(1998)}]{Ardell98}%
  \BibitemOpen
  \bibfield  {author} {\bibinfo {author} {\bibfnamefont {D.}~\bibnamefont {Ardell}},\ }\bibfield  {title} {\bibinfo {title} {On error minimization in a sequential origin of the standard genetic code},\ }\href {https://doi.org/10.1007/PL00006356} {\bibfield  {journal} {\bibinfo  {journal} {J Mol Evol}\ }\textbf {\bibinfo {volume} {47}},\ \bibinfo {pages} {1} (\bibinfo {year} {1998})}\BibitemShut {NoStop}%
\bibitem [{\citenamefont {Freeland}\ \emph {et~al.}(2000)\citenamefont {Freeland}, \citenamefont {Knight}, \citenamefont {Landweber},\ and\ \citenamefont {Hurst}}]{Freeland00}%
  \BibitemOpen
  \bibfield  {author} {\bibinfo {author} {\bibfnamefont {S.~J.}\ \bibnamefont {Freeland}}, \bibinfo {author} {\bibfnamefont {R.~D.}\ \bibnamefont {Knight}}, \bibinfo {author} {\bibfnamefont {L.~F.}\ \bibnamefont {Landweber}},\ and\ \bibinfo {author} {\bibfnamefont {L.~D.}\ \bibnamefont {Hurst}},\ }\bibfield  {title} {\bibinfo {title} {Early fixation of an optimal genetic code},\ }\href {https://doi.org/10.1007/PL00006356} {\bibfield  {journal} {\bibinfo  {journal} {Molecular Biology and Evolution}\ }\textbf {\bibinfo {volume} {17}},\ \bibinfo {pages} {511} (\bibinfo {year} {2000})}\BibitemShut {NoStop}%
\bibitem [{\citenamefont {Ardell}\ and\ \citenamefont {Sella}(2001)}]{Ardell2001}%
  \BibitemOpen
  \bibfield  {author} {\bibinfo {author} {\bibfnamefont {D.~H.}\ \bibnamefont {Ardell}}\ and\ \bibinfo {author} {\bibfnamefont {G.}~\bibnamefont {Sella}},\ }\bibfield  {title} {\bibinfo {title} {On the evolution of redundancy in genetic codes},\ }\href@noop {} {\bibfield  {journal} {\bibinfo  {journal} {Journal of molecular evolution}\ }\textbf {\bibinfo {volume} {53}},\ \bibinfo {pages} {269} (\bibinfo {year} {2001})}\BibitemShut {NoStop}%
\bibitem [{\citenamefont {Tlusty}(2008{\natexlab{a}})}]{Tlusty08_2}%
  \BibitemOpen
  \bibfield  {author} {\bibinfo {author} {\bibfnamefont {T.}~\bibnamefont {Tlusty}},\ }\bibfield  {title} {\bibinfo {title} {Rate-distortion scenario for the emergence and evolution of noisy molecular codes},\ }\href {https://doi.org/10.1103/PhysRevLett.100.048101} {\bibfield  {journal} {\bibinfo  {journal} {Physical review letters}\ }\textbf {\bibinfo {volume} {100}},\ \bibinfo {pages} {048101} (\bibinfo {year} {2008}{\natexlab{a}})}\BibitemShut {NoStop}%
\bibitem [{\citenamefont {Tlusty}(2008{\natexlab{b}})}]{Tlusty08}%
  \BibitemOpen
  \bibfield  {author} {\bibinfo {author} {\bibfnamefont {T.}~\bibnamefont {Tlusty}},\ }\bibfield  {title} {\bibinfo {title} {A simple model for the evolution of molecular codes driven by the interplay of accuracy, diversity and cost},\ }\href {https://doi.org/10.1088/1478-3975/5/1/016001} {\bibfield  {journal} {\bibinfo  {journal} {Physical biology}\ }\textbf {\bibinfo {volume} {5}},\ \bibinfo {pages} {016001} (\bibinfo {year} {2008}{\natexlab{b}})}\BibitemShut {NoStop}%
\bibitem [{\citenamefont {Tlusty}(2008{\natexlab{c}})}]{Tlusty2008c}%
  \BibitemOpen
  \bibfield  {author} {\bibinfo {author} {\bibfnamefont {T.}~\bibnamefont {Tlusty}},\ }\bibfield  {title} {\bibinfo {title} {Casting polymer nets to optimize noisy molecular codes},\ }\href@noop {} {\bibfield  {journal} {\bibinfo  {journal} {Proceedings of the National Academy of Sciences}\ }\textbf {\bibinfo {volume} {105}},\ \bibinfo {pages} {8238} (\bibinfo {year} {2008}{\natexlab{c}})}\BibitemShut {NoStop}%
\bibitem [{\citenamefont {Gilis}(2001)}]{Gilis01}%
  \BibitemOpen
  \bibfield  {author} {\bibinfo {author} {\bibfnamefont {D.}~\bibnamefont {Gilis}},\ }\bibfield  {title} {\bibinfo {title} {Optimality of the genetic code with respect to protein stability and amino-acid frequencies},\ }\href {https://doi.org/10.1186/gb-2001-2-11-research0049} {\bibfield  {journal} {\bibinfo  {journal} {Genome Biology}\ }\textbf {\bibinfo {volume} {2}},\ \bibinfo {pages} {253} (\bibinfo {year} {2001})}\BibitemShut {NoStop}%
\bibitem [{\citenamefont {Duchêne}\ \emph {et~al.}(2015)\citenamefont {Duchêne}, \citenamefont {Ho},\ and\ \citenamefont {Holmes}}]{Duchene15}%
  \BibitemOpen
  \bibfield  {author} {\bibinfo {author} {\bibfnamefont {S.}~\bibnamefont {Duchêne}}, \bibinfo {author} {\bibfnamefont {S.~Y.}\ \bibnamefont {Ho}},\ and\ \bibinfo {author} {\bibfnamefont {E.~C.}\ \bibnamefont {Holmes}},\ }\bibfield  {title} {\bibinfo {title} {Declining transition/transversion ratios through time reveal limitations to the accuracy of nucleotide substitution models},\ }\href {https://doi.org/10.1186/s12862-015-0312-6} {\bibfield  {journal} {\bibinfo  {journal} {BMC Evol Biol}\ }\textbf {\bibinfo {volume} {15}},\ \bibinfo {pages} {36} (\bibinfo {year} {2015})}\BibitemShut {NoStop}%
\bibitem [{\citenamefont {Begun}\ \emph {et~al.}(2007)\citenamefont {Begun}, \citenamefont {Holloway}, \citenamefont {Stevens}, \citenamefont {Hillier}, \citenamefont {Poh}, \citenamefont {Hahn}, \citenamefont {Nista}, \citenamefont {Jones}, \citenamefont {Kern}, \citenamefont {Dewey}, \citenamefont {Pachter}, \citenamefont {Myers},\ and\ \citenamefont {Langley}}]{Begun07}%
  \BibitemOpen
  \bibfield  {author} {\bibinfo {author} {\bibfnamefont {D.~J.}\ \bibnamefont {Begun}}, \bibinfo {author} {\bibfnamefont {A.~K.}\ \bibnamefont {Holloway}}, \bibinfo {author} {\bibfnamefont {K.}~\bibnamefont {Stevens}}, \bibinfo {author} {\bibfnamefont {L.~W.}\ \bibnamefont {Hillier}}, \bibinfo {author} {\bibfnamefont {Y.~P.}\ \bibnamefont {Poh}}, \bibinfo {author} {\bibfnamefont {M.~W.}\ \bibnamefont {Hahn}}, \bibinfo {author} {\bibfnamefont {P.~M.}\ \bibnamefont {Nista}}, \bibinfo {author} {\bibfnamefont {C.~D.}\ \bibnamefont {Jones}}, \bibinfo {author} {\bibfnamefont {A.~D.}\ \bibnamefont {Kern}}, \bibinfo {author} {\bibfnamefont {C.~N.}\ \bibnamefont {Dewey}}, \bibinfo {author} {\bibfnamefont {L.}~\bibnamefont {Pachter}}, \bibinfo {author} {\bibfnamefont {E.}~\bibnamefont {Myers}},\ and\ \bibinfo {author} {\bibfnamefont {C.~H.}\ \bibnamefont {Langley}},\ }\bibfield  {title} {\bibinfo {title} {Population genomics: whole-genome analysis of polymorphism and divergence in drosophila simulans},\ }\href
  {https://doi.org/10.1371/journal.pbio.0050310} {\bibfield  {journal} {\bibinfo  {journal} {PLoS biology}\ }\textbf {\bibinfo {volume} {5}},\ \bibinfo {pages} {e310} (\bibinfo {year} {2007})}\BibitemShut {NoStop}%
\bibitem [{\citenamefont {Hodgkinson}\ and\ \citenamefont {Eyre-Walker}(2010)}]{Hodgkinson10}%
  \BibitemOpen
  \bibfield  {author} {\bibinfo {author} {\bibfnamefont {A.}~\bibnamefont {Hodgkinson}}\ and\ \bibinfo {author} {\bibfnamefont {A.}~\bibnamefont {Eyre-Walker}},\ }\bibfield  {title} {\bibinfo {title} {Human triallelic sites: evidence for a new mutational mechanism?},\ }\href {https://doi.org/10.1534/genetics.109.110510} {\bibfield  {journal} {\bibinfo  {journal} {Genetics}\ }\textbf {\bibinfo {volume} {184}},\ \bibinfo {pages} {233} (\bibinfo {year} {2010})}\BibitemShut {NoStop}%
\bibitem [{\citenamefont {Jr.}(2019)}]{Saier19}%
  \BibitemOpen
  \bibfield  {author} {\bibinfo {author} {\bibfnamefont {M.~H.~S.}\ \bibnamefont {Jr.}},\ }\bibfield  {title} {\bibinfo {title} {Understanding the genetic code},\ }\href {https://doi.org/10.1128/jb.00091-19} {\bibfield  {journal} {\bibinfo  {journal} {Journal of bacteriology}\ }\textbf {\bibinfo {volume} {201}},\ \bibinfo {pages} {e00091} (\bibinfo {year} {2019})}\BibitemShut {NoStop}%
\bibitem [{\citenamefont {Goldman}(1993)}]{Goldman93}%
  \BibitemOpen
  \bibfield  {author} {\bibinfo {author} {\bibfnamefont {N.}~\bibnamefont {Goldman}},\ }\bibfield  {title} {\bibinfo {title} {Further results on error minimization in the genetic code},\ }\href {https://doi.org/10.1007/BF00182752} {\bibfield  {journal} {\bibinfo  {journal} {Journal of molecular evolution}\ }\textbf {\bibinfo {volume} {37}},\ \bibinfo {pages} {662} (\bibinfo {year} {1993})}\BibitemShut {NoStop}%
\bibitem [{\citenamefont {Giulio}(1989)}]{DiGiulio89}%
  \BibitemOpen
  \bibfield  {author} {\bibinfo {author} {\bibfnamefont {M.~D.}\ \bibnamefont {Giulio}},\ }\bibfield  {title} {\bibinfo {title} {The extension reached by the minimization of the polarity distances during the evolution of the genetic code},\ }\href {https://doi.org/10.1007/BF02103616} {\bibfield  {journal} {\bibinfo  {journal} {Journal of molecular evolution}\ }\textbf {\bibinfo {volume} {29}},\ \bibinfo {pages} {288} (\bibinfo {year} {1989})}\BibitemShut {NoStop}%
\bibitem [{\citenamefont {Mathew}\ and\ \citenamefont {Luthey-Schulten}(2008)}]{Mathew08}%
  \BibitemOpen
  \bibfield  {author} {\bibinfo {author} {\bibfnamefont {D.~C.}\ \bibnamefont {Mathew}}\ and\ \bibinfo {author} {\bibfnamefont {Z.}~\bibnamefont {Luthey-Schulten}},\ }\bibfield  {title} {\bibinfo {title} {On the physical basis of the amino acid polar requirement},\ }\href {https://doi.org/10.1007/s00239-008-9073-9} {\bibfield  {journal} {\bibinfo  {journal} {Journal of molecular evolution}\ }\textbf {\bibinfo {volume} {66}},\ \bibinfo {pages} {519} (\bibinfo {year} {2008})}\BibitemShut {NoStop}%
\bibitem [{\citenamefont {Grantham}(1974)}]{Grantham74}%
  \BibitemOpen
  \bibfield  {author} {\bibinfo {author} {\bibfnamefont {R.}~\bibnamefont {Grantham}},\ }\bibfield  {title} {\bibinfo {title} {Amino acid difference formula to help explain protein evolution},\ }\href {DOI: 10.1126/science.185.4154.862} {\bibfield  {journal} {\bibinfo  {journal} {Science}\ }\textbf {\bibinfo {volume} {185}},\ \bibinfo {pages} {862} (\bibinfo {year} {1974})}\BibitemShut {NoStop}%
\bibitem [{\citenamefont {Miyata}\ \emph {et~al.}(1979)\citenamefont {Miyata}, \citenamefont {Miyazawa},\ and\ \citenamefont {Yasunaga}}]{Miyata79}%
  \BibitemOpen
  \bibfield  {author} {\bibinfo {author} {\bibfnamefont {T.}~\bibnamefont {Miyata}}, \bibinfo {author} {\bibfnamefont {S.}~\bibnamefont {Miyazawa}},\ and\ \bibinfo {author} {\bibfnamefont {T.}~\bibnamefont {Yasunaga}},\ }\bibfield  {title} {\bibinfo {title} {Two types of amino acid substitutions in protein evolution},\ }\href {https://doi.org/10.1007/BF01732340} {\bibfield  {journal} {\bibinfo  {journal} {Journal of molecular evolution}\ }\textbf {\bibinfo {volume} {12}},\ \bibinfo {pages} {219} (\bibinfo {year} {1979})}\BibitemShut {NoStop}%
\bibitem [{\citenamefont {Rao}(1987)}]{Rao87}%
  \BibitemOpen
  \bibfield  {author} {\bibinfo {author} {\bibfnamefont {J.~K.~M.}\ \bibnamefont {Rao}},\ }\bibfield  {title} {\bibinfo {title} {New scoring matrix for amino acid residue exchanges based on residue characteristic physical parameters},\ }\href {https://doi.org/10.1111/j.1399-3011.1987.tb02254.x} {\bibfield  {journal} {\bibinfo  {journal} {International journal of peptide and protein research}\ }\textbf {\bibinfo {volume} {29}},\ \bibinfo {pages} {276} (\bibinfo {year} {1987})}\BibitemShut {NoStop}%
\bibitem [{\citenamefont {Prlić}\ \emph {et~al.}(2000)\citenamefont {Prlić}, \citenamefont {Domingues},\ and\ \citenamefont {Sippl}}]{Prlic00}%
  \BibitemOpen
  \bibfield  {author} {\bibinfo {author} {\bibfnamefont {A.}~\bibnamefont {Prlić}}, \bibinfo {author} {\bibfnamefont {F.~S.}\ \bibnamefont {Domingues}},\ and\ \bibinfo {author} {\bibfnamefont {M.~J.}\ \bibnamefont {Sippl}},\ }\bibfield  {title} {\bibinfo {title} {Structure-derived substitution matrices for alignment of distantly related sequences},\ }\href {https://doi.org/10.1093/protein/13.8.545} {\bibfield  {journal} {\bibinfo  {journal} {Protein engineering}\ }\textbf {\bibinfo {volume} {13}},\ \bibinfo {pages} {545} (\bibinfo {year} {2000})}\BibitemShut {NoStop}%
\bibitem [{\citenamefont {Yampolsky}\ and\ \citenamefont {Stoltzfus}(2005)}]{Yampolsky05}%
  \BibitemOpen
  \bibfield  {author} {\bibinfo {author} {\bibfnamefont {L.~Y.}\ \bibnamefont {Yampolsky}}\ and\ \bibinfo {author} {\bibfnamefont {A.}~\bibnamefont {Stoltzfus}},\ }\bibfield  {title} {\bibinfo {title} {The exchangeability of amino acids in proteins},\ }\href {https://doi.org/10.1534/genetics.104.039107} {\bibfield  {journal} {\bibinfo  {journal} {Genetics}\ }\textbf {\bibinfo {volume} {170}},\ \bibinfo {pages} {1459} (\bibinfo {year} {2005})}\BibitemShut {NoStop}%
\bibitem [{\citenamefont {Dayhoff}\ \emph {et~al.}(1978{\natexlab{a}})\citenamefont {Dayhoff}, \citenamefont {Schwartz},\ and\ \citenamefont {Orcutt}}]{Dayhoff78}%
  \BibitemOpen
  \bibfield  {author} {\bibinfo {author} {\bibfnamefont {M.~O.}\ \bibnamefont {Dayhoff}}, \bibinfo {author} {\bibfnamefont {R.~M.}\ \bibnamefont {Schwartz}},\ and\ \bibinfo {author} {\bibfnamefont {B.~C.}\ \bibnamefont {Orcutt}},\ }\bibfield  {title} {\bibinfo {title} {A model of evolutionary change in proteins},\ }in\ \href@noop {} {\emph {\bibinfo {booktitle} {Atlas of Protein Sequence and Structure}}},\ Vol.~\bibinfo {volume} {5}\ (\bibinfo  {publisher} {Natl. Biomed. Res.},\ \bibinfo {address} {Washington, D. C.},\ \bibinfo {year} {1978})\ \bibinfo {type} {Chapter}~\bibinfo {chapter} {22}, pp.\ \bibinfo {pages} {345--352}\BibitemShut {NoStop}%
\bibitem [{\citenamefont {Massey}(2016)}]{Massey16}%
  \BibitemOpen
  \bibfield  {author} {\bibinfo {author} {\bibfnamefont {S.~E.}\ \bibnamefont {Massey}},\ }\bibfield  {title} {\bibinfo {title} {The neutral emergence of error minimized genetic codes superior to the standard genetic code},\ }\href {https://doi.org/10.1016/j.jtbi.2016.08.022} {\bibfield  {journal} {\bibinfo  {journal} {Journal of theoretical biology}\ }\textbf {\bibinfo {volume} {408}},\ \bibinfo {pages} {237} (\bibinfo {year} {2016})}\BibitemShut {NoStop}%
\bibitem [{\citenamefont {Novozhilov}\ \emph {et~al.}(2007)\citenamefont {Novozhilov}, \citenamefont {Wolf},\ and\ \citenamefont {Koonin}}]{Novozhilov07}%
  \BibitemOpen
  \bibfield  {author} {\bibinfo {author} {\bibfnamefont {A.~S.}\ \bibnamefont {Novozhilov}}, \bibinfo {author} {\bibfnamefont {Y.~I.}\ \bibnamefont {Wolf}},\ and\ \bibinfo {author} {\bibfnamefont {E.~V.}\ \bibnamefont {Koonin}},\ }\bibfield  {title} {\bibinfo {title} {Evolution of the genetic code: partial optimization of a random code for robustness to translation error in a rugged fitness landscape},\ }\href {https://doi.org/10.1186/1745-6150-2-24} {\bibfield  {journal} {\bibinfo  {journal} {Biology Direct}\ }\textbf {\bibinfo {volume} {2}},\ \bibinfo {pages} {24} (\bibinfo {year} {2007})}\BibitemShut {NoStop}%
\bibitem [{\citenamefont {Choi}\ \emph {et~al.}(2017)\citenamefont {Choi}, \citenamefont {Gilson},\ and\ \citenamefont {Shakhnovich}}]{Choi17}%
  \BibitemOpen
  \bibfield  {author} {\bibinfo {author} {\bibfnamefont {J.~M.}\ \bibnamefont {Choi}}, \bibinfo {author} {\bibfnamefont {A.~I.}\ \bibnamefont {Gilson}},\ and\ \bibinfo {author} {\bibfnamefont {E.~I.}\ \bibnamefont {Shakhnovich}},\ }\bibfield  {title} {\bibinfo {title} {Graph's topology and free energy of a spin model on the graph},\ }\href {https://doi.org/10.1103/PhysRevLett.118.088302} {\bibfield  {journal} {\bibinfo  {journal} {Physical review letters}\ }\textbf {\bibinfo {volume} {118}},\ \bibinfo {pages} {088302} (\bibinfo {year} {2017})}\BibitemShut {NoStop}%
\bibitem [{\citenamefont {Dayhoff}\ and\ \citenamefont {Eck}(1969)}]{Dayhoff69}%
  \BibitemOpen
  \bibfield  {author} {\bibinfo {author} {\bibfnamefont {M.~O.}\ \bibnamefont {Dayhoff}}\ and\ \bibinfo {author} {\bibfnamefont {R.~V.}\ \bibnamefont {Eck}},\ }\href@noop {} {\emph {\bibinfo {title} {Atlas of Protein Sequence and Structure}}},\ Vol.~\bibinfo {volume} {2}\ (\bibinfo  {publisher} {Natl. Biomed. Res.},\ \bibinfo {address} {Silver Spring Md},\ \bibinfo {year} {1969})\BibitemShut {NoStop}%
\bibitem [{\citenamefont {Brooks}\ \emph {et~al.}(2002)\citenamefont {Brooks}, \citenamefont {Fresco}, \citenamefont {Lesk},\ and\ \citenamefont {Singh}}]{Brooks02}%
  \BibitemOpen
  \bibfield  {author} {\bibinfo {author} {\bibfnamefont {D.~J.}\ \bibnamefont {Brooks}}, \bibinfo {author} {\bibfnamefont {J.~R.}\ \bibnamefont {Fresco}}, \bibinfo {author} {\bibfnamefont {A.~M.}\ \bibnamefont {Lesk}},\ and\ \bibinfo {author} {\bibfnamefont {M.}~\bibnamefont {Singh}},\ }\bibfield  {title} {\bibinfo {title} {Evolution of amino acid frequencies in proteins over deep time: Inferred order of introduction of amino acids into the genetic code},\ }\href {https://doi.org/10.1093/oxfordjournals.molbev.a003988} {\bibfield  {journal} {\bibinfo  {journal} {Molecular biology and evolution}\ }\textbf {\bibinfo {volume} {19}},\ \bibinfo {pages} {1645} (\bibinfo {year} {2002})}\BibitemShut {NoStop}%
\bibitem [{\citenamefont {Jungck}(1978)}]{Jungck78}%
  \BibitemOpen
  \bibfield  {author} {\bibinfo {author} {\bibfnamefont {J.~R.}\ \bibnamefont {Jungck}},\ }\bibfield  {title} {\bibinfo {title} {The genetic code as a periodic table},\ }\href {https://doi.org/10.1007/BF01734482} {\bibfield  {journal} {\bibinfo  {journal} {Journal of molecular evolution}\ }\textbf {\bibinfo {volume} {11}},\ \bibinfo {pages} {211} (\bibinfo {year} {1978})}\BibitemShut {NoStop}%
\bibitem [{\citenamefont {Jukes}\ \emph {et~al.}(1975)\citenamefont {Jukes}, \citenamefont {Holmquist},\ and\ \citenamefont {Moise}}]{Jukes75}%
  \BibitemOpen
  \bibfield  {author} {\bibinfo {author} {\bibfnamefont {T.~H.}\ \bibnamefont {Jukes}}, \bibinfo {author} {\bibfnamefont {R.}~\bibnamefont {Holmquist}},\ and\ \bibinfo {author} {\bibfnamefont {H.}~\bibnamefont {Moise}},\ }\bibfield  {title} {\bibinfo {title} {Amino acid composition of proteins: Selection against the genetic code},\ }\href {https://doi.org/10.1126/science.237322} {\bibfield  {journal} {\bibinfo  {journal} {Science}\ }\textbf {\bibinfo {volume} {189}},\ \bibinfo {pages} {50} (\bibinfo {year} {1975})}\BibitemShut {NoStop}%
\bibitem [{\citenamefont {Dayhoff}\ \emph {et~al.}(1978{\natexlab{b}})\citenamefont {Dayhoff}, \citenamefont {Hunt},\ and\ \citenamefont {Hurst-Calderone}}]{Dayhoff78_2}%
  \BibitemOpen
  \bibfield  {author} {\bibinfo {author} {\bibfnamefont {M.~O.}\ \bibnamefont {Dayhoff}}, \bibinfo {author} {\bibfnamefont {L.~T.}\ \bibnamefont {Hunt}},\ and\ \bibinfo {author} {\bibfnamefont {S.}~\bibnamefont {Hurst-Calderone}},\ }\href@noop {} {\emph {\bibinfo {title} {Atlas of Protein Sequence and Structure}}},\ Vol.~\bibinfo {volume} {5}\ (\bibinfo  {publisher} {Natl. Biomed. Res.},\ \bibinfo {address} {Washington, D. C.},\ \bibinfo {year} {1978})\BibitemShut {NoStop}%
\bibitem [{\citenamefont {Balin}\ and\ \citenamefont {Cascalho}(2010)}]{Balin10}%
  \BibitemOpen
  \bibfield  {author} {\bibinfo {author} {\bibfnamefont {S.~J.}\ \bibnamefont {Balin}}\ and\ \bibinfo {author} {\bibfnamefont {M.}~\bibnamefont {Cascalho}},\ }\bibfield  {title} {\bibinfo {title} {The rate of mutation of a single gene},\ }\href {https://doi.org/10.1093/nar/gkp1119} {\bibfield  {journal} {\bibinfo  {journal} {Nucleic acids research}\ }\textbf {\bibinfo {volume} {38}},\ \bibinfo {pages} {1575} (\bibinfo {year} {2010})}\BibitemShut {NoStop}%
\bibitem [{\citenamefont {Suzuki}\ and\ \citenamefont {Nagao}(2021)}]{Suzuki21}%
  \BibitemOpen
  \bibfield  {author} {\bibinfo {author} {\bibfnamefont {T.}~\bibnamefont {Suzuki}}\ and\ \bibinfo {author} {\bibfnamefont {A.}~\bibnamefont {Nagao}},\ }\bibinfo {title} {Genetic code and its variations},\ in\ \href {https://doi.org/10.1002/9780470015902.a0029263} {\emph {\bibinfo {booktitle} {eLS}}}\ (\bibinfo  {publisher} {John Wiley \& Sons, Ltd},\ \bibinfo {year} {2021})\ pp.\ \bibinfo {pages} {147--157}\BibitemShut {NoStop}%
\bibitem [{\citenamefont {Lajoie}\ \emph {et~al.}(2013)\citenamefont {Lajoie}, \citenamefont {Rovner}, \citenamefont {Goodman}, \citenamefont {Aerni}, \citenamefont {Haimovich}, \citenamefont {Kuznetsov}, \citenamefont {Mercer}, \citenamefont {Wang}, \citenamefont {Carr}, \citenamefont {Mosberg}, \citenamefont {Rohland}, \citenamefont {Schultz}, \citenamefont {Jacobson}, \citenamefont {Rinehart}, \citenamefont {Church},\ and\ \citenamefont {Isaacs}}]{Lajoie13}%
  \BibitemOpen
  \bibfield  {author} {\bibinfo {author} {\bibfnamefont {M.~J.}\ \bibnamefont {Lajoie}}, \bibinfo {author} {\bibfnamefont {A.~J.}\ \bibnamefont {Rovner}}, \bibinfo {author} {\bibfnamefont {D.~B.}\ \bibnamefont {Goodman}}, \bibinfo {author} {\bibfnamefont {H.~R.}\ \bibnamefont {Aerni}}, \bibinfo {author} {\bibfnamefont {A.~D.}\ \bibnamefont {Haimovich}}, \bibinfo {author} {\bibfnamefont {G.}~\bibnamefont {Kuznetsov}}, \bibinfo {author} {\bibfnamefont {J.~A.}\ \bibnamefont {Mercer}}, \bibinfo {author} {\bibfnamefont {H.~H.}\ \bibnamefont {Wang}}, \bibinfo {author} {\bibfnamefont {P.~A.}\ \bibnamefont {Carr}}, \bibinfo {author} {\bibfnamefont {J.~A.}\ \bibnamefont {Mosberg}}, \bibinfo {author} {\bibfnamefont {N.}~\bibnamefont {Rohland}}, \bibinfo {author} {\bibfnamefont {P.~G.}\ \bibnamefont {Schultz}}, \bibinfo {author} {\bibfnamefont {J.~M.}\ \bibnamefont {Jacobson}}, \bibinfo {author} {\bibfnamefont {J.}~\bibnamefont {Rinehart}}, \bibinfo {author} {\bibfnamefont {G.~M.}\ \bibnamefont {Church}},\ and\ \bibinfo
  {author} {\bibfnamefont {F.~J.}\ \bibnamefont {Isaacs}},\ }\bibfield  {title} {\bibinfo {title} {Genomically recoded organisms expand biological functions},\ }\href {https://doi.org/10.1126/science.1241459} {\bibfield  {journal} {\bibinfo  {journal} {Science}\ }\textbf {\bibinfo {volume} {342}},\ \bibinfo {pages} {357} (\bibinfo {year} {2013})}\BibitemShut {NoStop}%
\bibitem [{\citenamefont {Chin}(2017)}]{Chin17}%
  \BibitemOpen
  \bibfield  {author} {\bibinfo {author} {\bibfnamefont {J.}~\bibnamefont {Chin}},\ }\bibfield  {title} {\bibinfo {title} {Expanding and reprogramming the genetic code},\ }\href {https://doi.org/10.1038/nature24031} {\bibfield  {journal} {\bibinfo  {journal} {Nature}\ }\textbf {\bibinfo {volume} {550}},\ \bibinfo {pages} {53} (\bibinfo {year} {2017})}\BibitemShut {NoStop}%
\bibitem [{\citenamefont {Buhrman}\ \emph {et~al.}(2011)\citenamefont {Buhrman}, \citenamefont {van~der Gulik}, \citenamefont {Kelk}, \citenamefont {Koolen},\ and\ \citenamefont {Stougie}}]{Buhrman11}%
  \BibitemOpen
  \bibfield  {author} {\bibinfo {author} {\bibfnamefont {H.}~\bibnamefont {Buhrman}}, \bibinfo {author} {\bibfnamefont {P.~T.}\ \bibnamefont {van~der Gulik}}, \bibinfo {author} {\bibfnamefont {S.~M.}\ \bibnamefont {Kelk}}, \bibinfo {author} {\bibfnamefont {W.~M.}\ \bibnamefont {Koolen}},\ and\ \bibinfo {author} {\bibfnamefont {L.}~\bibnamefont {Stougie}},\ }\bibfield  {title} {\bibinfo {title} {Some mathematical refinements concerning error minimization in the genetic code},\ }\href {https://doi.org/10.1109/TCBB.2011.40} {\bibfield  {journal} {\bibinfo  {journal} {IEEE/ACM transactions on computational biology and bioinformatics}\ }\textbf {\bibinfo {volume} {8}},\ \bibinfo {pages} {1358} (\bibinfo {year} {2011})}\BibitemShut {NoStop}%
\end{thebibliography}%

\end{document}